\documentclass[a4paper,journal]{IEEEtran}
\IEEEoverridecommandlockouts
\usepackage{booktabs}
\usepackage{cite}
\usepackage{floatrow}
\usepackage{amsmath,amssymb,amsfonts}
\usepackage{color}
\usepackage{graphicx}
\usepackage{textcomp}
\usepackage{xcolor}
\usepackage{array}
\usepackage{subfigure}
\usepackage{subfigure}
\usepackage{mathtools, amssymb, amsthm}
\newtheorem{theorem}{\noindent\rm\textbf{Theorem}}
\newtheorem{definition}{\noindent\rm\textbf{Definition}}
\newtheorem{lemma}{\noindent\rm\textbf{Lemma}}
\newtheorem{assumption}{\rm\bf\noindent Assumption}

\newtheorem{proposition}{\noindent\rm\textbf{Proposition}}
\usepackage[justification=centering]{caption}
\usepackage[ruled,vlined]{algorithm2e}
\usepackage{algorithmic}
\usepackage{amsmath}
\usepackage{mathrsfs}
\usepackage{graphics}
\usepackage{epsfig}
\usepackage{url}

\usepackage{lineno}

\renewcommand{\algorithmicrequire}{\textbf{Input:}} 
\renewcommand{\algorithmicensure}{\textbf{Output:}} 

\begin{document}
\title{Privacy Preserving in Non-Intrusive Load Monitoring: A Differential Privacy Perspective}

\author{Haoxiang Wang, Jiasheng Zhang, Chenbei Lu, \emph{and} Chenye Wu \vspace{-0.8cm}
\thanks{H. Wang, J. Zhang and C. Lu are with the Institute for Interdisciplinary Information Sciences (IIIS), Tsinghua University. C. Wu is with School of Science and Engineering, The Chinese University of Hong Kong, Shenzhen. C. Wu is the correspondence author. Email: chenyewu@yeah.net.}
}

\maketitle

\begin{abstract}
Smart meter devices enable a better understanding of the demand at the potential risk of private information leakage. One promising solution to mitigating such risk is to inject noises into the meter data to achieve a certain level of differential privacy. In this paper, we cast one-shot non-intrusive load monitoring (NILM) in the compressive sensing framework, and bridge the gap between theoretical accuracy of NILM inference and differential privacy's parameters. We then derive the valid theoretical bounds to offer insights on how the differential privacy parameters affect the NILM performance. Moreover, we generalize our conclusions by proposing the hierarchical framework to solve the multi-shot NILM problem. Numerical experiments verify our analytical results and offer better physical insights of differential privacy in various practical scenarios. This also demonstrates the significance of our work for the general privacy preserving mechanism design. 

\end{abstract}


\section{Introduction}\label{sect1}
The pervasive intelligent devices are gathering {huge amounts of data} in our daily lives, spanning from our {shopping lists} to our favorite restaurants, from travel {histories} to personal social {networks}. These big data have eased our social lives and {have} dramatically changed our behaviors. In the electricity sector, the widely deployed smart meters are collecting user's energy consumption data in near real time. While these data could be rather valuable to achieve a more efficient power system, they raise significant public concern on private information leakage. Specifically, the big data in the electricity sector speed up the advance in non-intrusive load monitoring (NILM), which aims to infer the user's energy consumption pattern from meter data.

{NILM is one of the most effective ways to conduct consumer behavior analysis \cite{hart1989residential,kreith2017crc}. More comprehensive behavior analysis could benefit the consumers by providing ambient assisted living \cite{fell2017energising}, real time energy saving \cite{herrero2017non}, etc. It is evident that such analysis is crucial for the active demand side management, which is believed to be able to significantly improve the whole system efficiency \cite{31557}}. However, the inferred consumption pattern often exposes individual lifestyles.
This implies that the leakage of smart meter data may lead to {the concern over the leakage of private information,} 
which calls for a comprehensive privacy preservation scheme. The European Union is the pioneer in customer privacy protection:
it has legislated an anti-privacy data protection regulation in 2018 \cite{8283439}.
Brazil also {has also set} up the General Data Protection Law which become effective in February 2020 \cite{brazil.org}.
Moreover, 11 states in the US have recently enacted privacy, data security, cybersecurity, and data breach notification laws \cite{11usa.org}.

To achieve the privacy protection, the most commonly adopted technique is differential privacy (DP), first proposed by Dwork \emph{et al}. in \cite{dwork2006calibrating}. DP facilitates the mathematical analysis and is also closely related to other privacy metrics such as mutual information \cite{wang2016relation}. 
However, despite well investigated, the 
parameters in DP do not offer intuitive physical {insights}, which prevents this technique from wide deployment. Hence, it is a delicate task to design a practical privacy preserving mechanism of NILM.

In this work, we submit that DP indeed has physical implications in
the electricity sector. We propose to understand DP through
NILM and characterize 
how the parameters in DP affect the performance guarantee of 
NILM inference. Our work offers the end users a better idea on the different levels of privacy preserving services in gathering meter data.

\subsection{Related Works}

NILM and DP are both well investigated. 
Since the seminal work by Hart \cite{31557}, diverse techniques for NILM have been 
proposed to solve NILM for improved {inference} performances. Our work focuses on the 
disaggregation process in NILM. The classical algorithm is 
Combinatorial Optimization (CO), proposed in Hart's seminal work. This algorithm
combines heuristic methods with prior knowledge on the switching events.
Zulfiquar
\emph{et al}. {have furthered} this research by introducing the aided linear integer programming technique to speed up {CO} in \cite{bhotto2017load}.
Probabilistic methods have also been proposed recently. For example, Kim \emph{et al}. {have applied} Factorial Hidden Markov Model (FHMM) to NILM, and {have} used Viterbi algorithm for decoding \cite{kim2011unsupervised}. The key challenge {in} this line of research is to model the appropriate states {for} the hidden Markov model.
Makonin \emph{et al.} {have used} machine learning techniques to solve this issue and {have proposed} the notion of super-state HMM \cite{makonin2016exploiting}.
The advance in deep learning warrants {exploiting} more temporal features in NILM. For example,  Kelly \emph{et al.} {have deployed} the long short-time memory (LSTM) framework for NILM in \cite{kelly2015neural}.

There is a recent interest in designing the privacy preserving NILM, and our work also falls into this category. However, most literature {have focused} on discussing thwarting the privacy attacks on the meter data \cite{kalogridis2010privacy}. The major technique is to utilize the storage system to physically inject noises into the meter data, which does offer the end users a certain privacy guarantee. Backes \emph{et al.} {have theoretically examined} the way to achieve different levels of privacy using storage injection manipulation in \cite{backes2012differentially}. {More} recently, {a variety of} methods have been exploited to achieve privacy preservation.

Chen \emph{et al}. {
have utilized} the combined heat and privacy system to prevent
occupancy detection in \cite{chen2015preventing}. 
Cao \emph{et al}. {
have investigated} the practical implementation {from a fog computing approach} \cite{cao2019achieving}.
Rastogi \emph{et al}. {have further proposed} a distributed
implementation in \cite{rastogi2010differentially}.

{To the best of our knowledge, we are the first to exploit the physical meaning of privacy parameters in the electricity sector.}

\subsection{Our Contributions}\label{contribution}

In establishing the connection between DP parameters and the inference accuracy of NILM, our {principal} contributions can be summarized as follows:
\begin{itemize}
  \item
  \textit{NILM Formulation with DP}: 
  We use the compressive sensing framework to formulate NILM inference and incorporate the parameters of DP into the formulation. 
  \item \textit{Theoretical Characterization of NILM Inference}: Based on the compressive sensing formulation, we theoretically characterize the asymptotic upper and lower bounds for the one-shot NILM inference accuracy. 
  \item \textit{Hierarchical Multi-shot NILM}: We generalize the one-shot NILM solution to {more practical} multi-shot scenarios, and propose an effective hierarchical algorithm.
\end{itemize}

{We imagine our proposed framework has at least three early adopters.}

\begin{itemize}
    \item {The first adopter could be the consumers themselves. To preserve privacy, the consumers could utilize the local storage devices (provided by electric vehicles, photovoltaic panels, etc.) to inject noises to achieve certain level of differential privacy. Our theoretical understanding provides an inference accuracy bound to decipher the privacy preserving guarantee.}
    \item {The other adopter could be the ISOs or the utility companies. When recording the meter data from the consumers for potential behavior analysis, the ISOs or utility companies could directly inject noises into the recorded meter data. Due to the Law of Larger Numbers, the injected noise will incur minimal impact on auditing (in fact, auditing before injecting the noise could solve this issue) and most classical tasks of utility companies. However, the injected noise could affect behavior analysis (see detailed discussion in Appendix \ref{inflence_and_implement}). This is another way to decipher the physical meaning of DP parameters.}
    \item {The last adopter could be the third-party privacy preserving entities. They could collect the data from the consumers for different purposes. And for different privacy preserving requirements, the consumers are paid for different compensations. This could enable new business models for the privacy preserving industry, and our theoretical results could give the first cut understanding on the cost benefit analysis for the emerging business models.}
\end{itemize}

\begin{figure}[!t]
  \centering
  \vspace{0.2cm}
  \includegraphics[width=3. in]{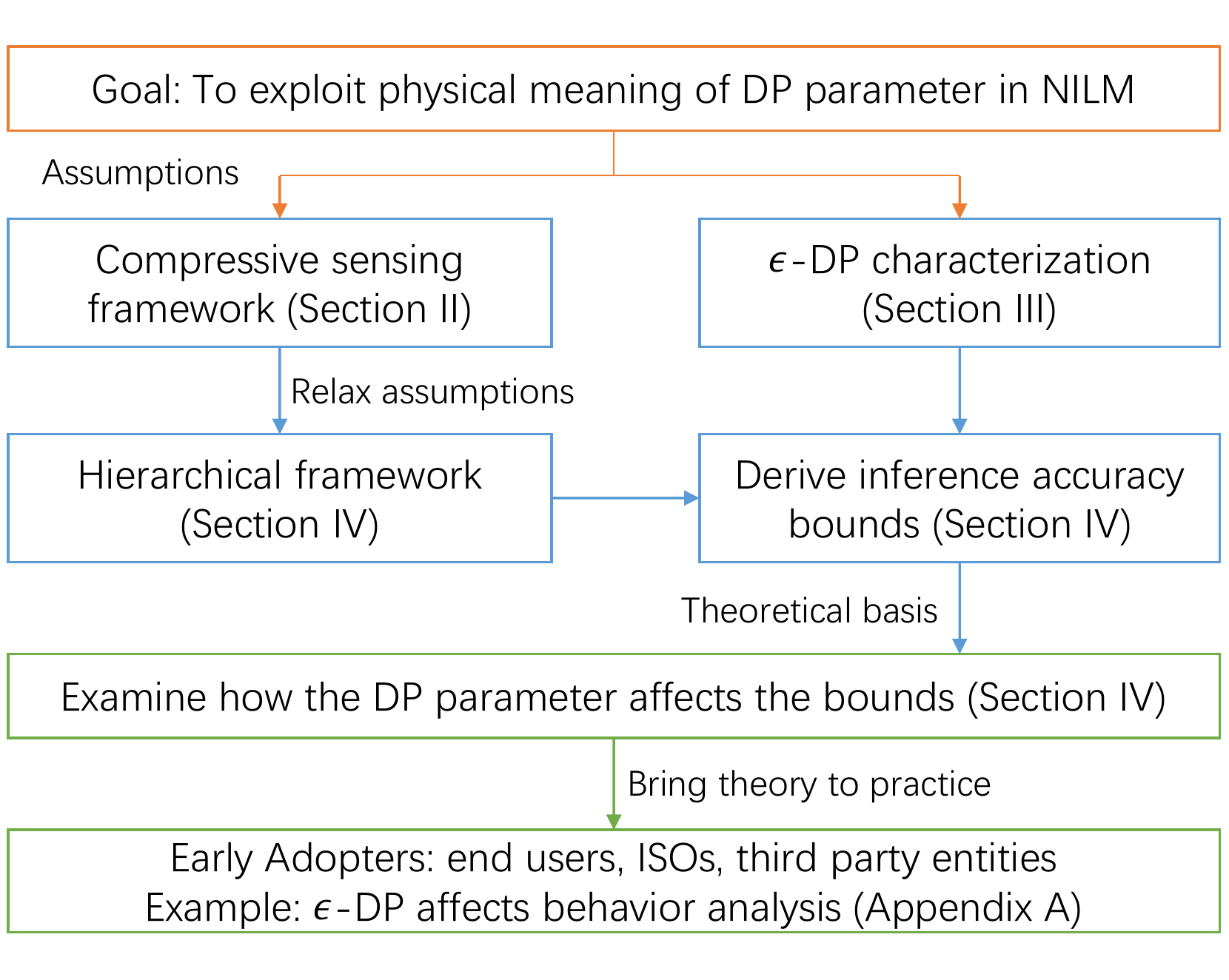}\vspace{-0.2cm}
  \caption{The paradigm of our paper}
  \label{paradigm}
  \end{figure}

The rest of the paper is organized as follows. 
Section \ref{NILM} formulates NILM as a sparse 
optimization problem. Using the compressive sensing 
framework, we establish the lower and upper
bounds of inference accuracy for NILM, and relate the 
$\epsilon$-differential privacy to these asymptotic bounds
in Section \ref{DP} to better understand the physical 
implication of DP. Then, in Section \ref{muti}, we generalize the compressive sensing framework to the multi-shot scenario and introduce our hierarchical algorithmic framework. Numerical studies verify the effectiveness of our theoretical {conclusion} in Section \ref{experiments}. Concluding remarks and several interesting future directions are given in Section 
\ref{conclusion}. We provide all the necessary proofs and {justifications} in Appendix. {We visualize the paradigm of our paper in Fig. \ref{paradigm}, which highlights the underlying logic between sections.}

\section{Problem Formulation}\label{NILM}

This section first introduces the general NILM problem, and then formulates the one-shot NILM in the compressive sensing framework. We conclude by revisiting the generic compressive sensing algorithm to solve the resulting NILM problem.

\subsection{Non-intrusive Load Monitoring Problem}\label{problem formulation}
In essence, the goal of NILM is to infer user behaviors from meter data. Mathematically, 
for each end user (or a building complex, a campus, etc.), it may own $N$ 
appliances, denoted by a set $\mathcal{A}=\{A_{1},..,A_{N}\}$.

{For simplicity, we assume the state space for each appliance is binary, i.e., on or off.}
For appliance $i\in \mathcal{A}$, {when its state is on}, its energy consumption at each time $t$ 
is a random variable $Z_{i}^{t}$, with mean of $P_{i}$. Define {the vector 
of mean energy consumption over all appliances} $\mathbf{P}$ as follows:
\begin{equation}
\begin{aligned}
\label{eq1}
\mathbf{P}=(P_{1},..,P_{N})^\dag \in \mathbf{R}^{N}.
\end{aligned}
\end{equation}

Hence, at each time $t$, the energy consumption of the end user is this
aggregate energy consumption of {its appliances whose states are on}. {We denote this} subset 
by $S_{t} \subseteq \{1,...,N\}$, and denote the meter data at time $t$ by $y_{t}$. 
Thus, by definition of $y_t$, we know 
\begin{equation}
\begin{aligned}\label{y_t}
y_{t}=\sum\nolimits_{i\in S_{t}}Z_{i}^{t},\,\,\,\,\,\,\,\  \forall t=1,...,T,
\end{aligned}
\end{equation}
where $T$ stands for the length of observation period. If we consider 
the end user's hourly operation with a resolution of $6$ seconds, then $T$ is $600$.
The advanced metering technology has made it possible to collect data at much finer resolution, e.g., in a sub-second scale \cite{makonin2018rae:}. Such data could allow us to exploit more {unique} energy consumption patterns for each appliance. However, in this paper, we choose to focus on a stylized model to better clarify the connection between DP and NILM, without specifying the resolution.

The classical NILM seeks to infer the on/off switch events of all appliances. We denote $X_{t} \in \{0,1\}^{N}$ the on/off indicator of the $N$ 
appliances at time $t$ ({with zero indicating off and one indicating on}). NILM {aims at} inferring $X_{t}$ from $y_{t}$.

A variety of NILM algorithms have been proposed as we have reviewed in Section \ref{sect1}. In this work, we choose the compressive sensing framework for {mathematically} exploiting the
sparsity structure in the event vector $X_t$. However, conducting NILM inference can be delicate as the energy consumptions in $\mathcal{A}$ can be quite diverse, and some `elephant' appliances (that consume a huge amount of power) {may} dominate the period of interest. In this case, it is almost impossible to accurately distinguish the on/off states of small appliances. In the following formulation, we devise a sufficient condition to enable accurate NILM inference, {based on which, we propose the hierarchical framework for more practical situations in the subsequence analysis.}

\subsection{One-shot NILM Inference}
To better characterize the relationship between
differential privacy and NILM, 
we focus on the inference for a 
particular time $t$ and assume the knowledge 
of $X_{t-1}$. We term this problem 
the \emph{one-shot NILM inference}.

{To exploit the sparse structure,} the 
compressive sensing formulation {requires the following transforms:} 
\begin{equation}
\begin{aligned}\label{K_t}
K_{t} = |y_{t}-y_{t-1}| \in \mathbf{R},
\end{aligned}
\end{equation}
\begin{equation}
\begin{aligned}\label{delta_t}
\Delta_{t}=|X_{t}-X_{t-1}| \in \mathbf{R}^{N}.
\end{aligned}
\end{equation}
With these transforms, we can mathematically characterize the sparsity assumption.

\vspace{0.1cm}
\begin{assumption}\label{assumption1}\rm
 (Sparsity Assumption) The switch events are sparse 
{across time}. That is, the number of switch events in $\Delta_{t}$ (i.e., $\|\Delta_t\|_0$) is bounded by $U_{t}$, which satisfies 
\begin{align}
    \|\Delta_{t}\|_{0}\leq U_{t}\ll{N}.
\end{align}
\end{assumption}

\vspace{0.1cm}
\noindent {$\textbf{Remark:}$ This assumption holds for most publicly available datasets with the resolution on the second scale. In Appendix \ref{sparsity_sec}, we characterize the sparsity of three most widely adopted public datasets and justify how the failure to meet this assumption may affect the performance of our proposed framework.
}

This sparsity assumption motivates us to formulate the following optimization problem for the one-shot NILM inference at each time $t$:
\begin{equation}\label{The problem P1}
\begin{aligned}
(\text{P1})\,\,\ \min\,\,&\|\Delta_{t}\|_{0}  \\
s.t.\,\,\,\,&\|\Delta_{t}\mathbf{P}-K_{t}\|_{2}<\delta,\\
&  \Delta_{t}\in \{0,1\}^{N},
\end{aligned}
\end{equation}
where $\delta$ is the parameter to characterize the sensitivity of $y_t$.

The notion of sensitivity is defined for random variables:

\vspace{0.1cm}
\begin{definition}\label{def1}\rm
 We define $\Delta f$ to be the sensitivity of {a sequence of bounded random variable} $y_{t}$:
 \begin{equation}
\label{dltf}
\Delta f=\max_{1\le t\le T}\left|\bar{y}_{t}-\underline{y}_{t}\right|=\frac{\delta}{2},
\end{equation}
where $\underline{y}_{t}$ and $\bar{y}_{t}$ are the lower and upper bounds of $y_t$.
\vspace{0.1cm}
\end{definition}

Back to our problem (P1), the sensitivity constraints require the meter data are bounded, {which yields our second assumption}:

\vspace{0.1cm}
\begin{assumption}\label{assumption2}\rm
 The meter data\footnote{{We only consider the pure load meter data in this work. To handle the net metering cases, we need to first resort to load disaggregation techniques \cite{kara2018disaggregating,tabone2018disaggregating,wang2018distributed}, and then apply our framework to the disaggregation pure load data.}} $y_{t}$ is bounded, i.e.,
 \begin{align}
    y_{t}\in \left[\underline{y}_{t} , \bar{y}_{t}\right], 
    1\le t\le T.
\end{align}
\end{assumption}

This assumption is {straightforward} to justify: it simply suggests that when the appliances are running, their {energy consumption levels are} within the specific range. And {such} range corresponds to the sensitivity parameter $\delta$.

\vspace{0.1cm}
\noindent {\textbf{Remark:} This assumption also partially handles the measurement error. Such error could be due to environmental issues, operational issues, or other issues. Practically, such error is not a major concern. The reason is two-folded. Firstly, as indicated in \cite{national2010ansi}, the magnitude of such error is required to be bounded by $5\%$ of the total power. Secondly, if the data quality is too poor to infer any useful information, there won't be any privacy preserving requirement at all.
}

Note that, (P1) is similar to the {classical} problem in the compressive sensing framework \cite{donoho2003optimally} but not exactly the same. We denote the optimal solution to (P1) by $\Delta_t^0$ and regard it as the ground truth. Deciphering this ground truth is quite challenging: the binary constraints {in addition to} the $l_0$-norm objective function make the problem intractable.

To tackle these challenges, we propose to first relax the binary constraints and then use the standard compressive sensing technique to handle the non-convexity induced by the $l_0$-norm objective function.

Specifically, we make the following technique alignment assumption to enable exact relaxation:

\vspace{0.1cm}
\begin{assumption}\label{assumption3}\rm(Power Concentration Assumption)
We assume the mean energy consumptions of
all the $N$ appliances {are on the same order}. Mathematically,
denote $\{P_{1}^\epsilon,...,P_{N}^\epsilon\}$ the ascendingly 
ordered sequence of $P$, we assume the following condition holds for 
all $U<U_{t}$:
\begin{equation}
\begin{aligned}
\sum\nolimits_{k=1}^{U}P_{k}^\epsilon-\sum\nolimits_{k=1}^{U-1}P_{N+1-k}^\epsilon>2\delta.
\end{aligned} 
\end{equation}
\end{assumption}

\vspace{0.1cm}
\noindent {$\textbf{Remark}$: Assumption 3 guarantees the effectiveness of the compressive sensing framework. This condition tries to eliminate the task to distinguish the lightning load from the refrigerator or the heating load, which allows us to focus on the inference of loads of similar energy consumption levels.  
For cases with diverse energy consumptions, we defer the detailed discussion to Section \ref{hmutialgo}. The general idea is to design a hierarchical NILM, where in each hierarchy, the NILM inference satisfies this technical alignment assumption.}

This assumption allows us to relax the binary constraints without changing the structure of the optimal solution set:
\begin{equation}\label{The problem P2}
\begin{aligned}
(\text{P2})\,\,\ \min\,\,&\|\Delta_{t}\|_{0}  \\
s.t.\,\,\,\,&\|\Delta_{t}\mathbf{P}-K_{t}\|_{2}<\delta,\\
&  \Delta_{t}\in [0,1]^{N}.
\end{aligned}
\end{equation}

We formally state the equivalence between (P1) and (P2) in the following Lemma.

\vspace{0.1cm}
\begin{lemma}\label{lemma1}\rm
If Assumptions 1 to 3 hold, problems (P1) and (P2) are equivalent in terms of the optimal solution set. Specifically, if we denote the optimal solution to (P1) by
$\Delta_{t}^{0}$ and the optimal solution to (P2) by $\Delta_{t}$, the equivalence is indicated by the following equation:
\begin{equation}
\| \Delta_t^0 \|_0 = \| \Delta_t \|_0.
\end{equation}
\end{lemma}

We show detailed proof in Appendix \ref{Lemma 1}. This lemma allows us to focus on solving a more tractable problem, (P2). {This is due to the key theoretical basis of compressive sensing:} we could effectively use $l_{1}$-norm 
to approximate $l_{0}$-norm in (P2) \cite{candes2006stable}, which yields (P3),
\begin{equation}\label{The problem P3}
\begin{aligned}
(\text{P3})\,\,\ \min\,\,&\|\Delta_{t}\|_{1}  \\
s.t.\,\,\,\,&\|\Delta_{t}\mathbf{P}-K_{t}\|_{2}<\delta,\\
&  \Delta_{t}\in [0,1]^{N}.
\end{aligned}
\end{equation}

Denote its optimal solution by $\Delta_{t}^{*}$. 
To map $\Delta_{t}^{*}$ onto $\{0,1\}^{N}$, we conduct rounding. 
Denote the solution after rounding by $\bar{\Delta}_{t}$. 
The rounding process is as follows: we first set all elements 
in $\bar{\Delta}_{t}$ to be $0$; and then, for each non-zero 
element $\Delta_{t}^{*}(j)$ in $\Delta_{t}^{*}$, with probability 
$\Delta_{t}^{*}(j)$, we set the corresponding element $\bar{\Delta}_{t}(j)$ to 
be $1$. To characterize the inaccuracy induced by the approximation of $l_{0}$-norm, 
we compare $\mathbf{E}[\bar{\Delta}_{t}]$ 
and the ground truth, $\Delta_{t}^{0}$. The comparison yields the following theorem:

\vspace{0.1cm}
\begin{theorem}\label{theorem1_}\rm
On expectation, the difference between $\bar{\Delta}_{t}$ and $\Delta_{t}^{0}$ is bounded. Specifically, 
\begin{equation}\label{theorem1_delta}
   \|\mathbf{E}[\bar{\Delta}_t]-\Delta_{t}^{0}\|_{2}\leq C(\mathbf{P})\cdot\delta,
\end{equation}
where $C(\mathbf{P})$ is a constant uniquely determined by vector $\mathbf{P}$.
\vspace{0.1cm}
\end{theorem}
We characterize {the} specific form {of $C(\mathbf{P})$} {in} Appendix \ref{Proposition 1}.
To prove Theorem 1, we mainly make use of the important conclusion in compressive sensing proposed by Candes \emph{et al.}  in \cite{candes2006stable}.

\section{Encode DP into NILM}\label{DP}
Based on the classical compressive sensing methods to solve the one-shot NILM inference, we establish the connection between DP and NILM for exploiting the physical meaning of different DP parameters in the context of NILM.

\subsection{Revisit $\epsilon$-Differential Privacy}
We adopt the notion of $\epsilon$-differential privacy
\cite{dwork2006calibrating} to connect NILM with DP. It uses the parameter $\epsilon$ to denote the probability
that we {are unable to} differentiate the two datasets that have only one piece of data difference.

To put it more formally, we define a mapping $\mathscr{B}(D)$ from a dataset $D$ to $\mathbb{R}$. This mapping enables a query function $q:D\to\mathbb{R}$. To measure the difference between two datasets $D$ and $D'$, we adopt the distance metric $d(D,D')$, measuring the minimal number of sample changes that are required to make $D$ identical to $D'$. When $d(D,D') = 1$, we say $D$ and $D'$ are neighbor datasets. This allows us to characterize the notion of $\epsilon$-DP:
if for all neighbor datasets $D_1$ and $D_2$, and for all measurable subsets $Y \subset \mathbb{R}$, the mapping $\mathscr{B}$ satisfies,
\begin{equation}
    \frac{Pr(\mathscr{B}(D_1)\in Y)}{Pr(\mathscr{B}(D_2)\in Y)}\leq e^{\epsilon},
\end{equation}
we say the mechanism $\mathscr{B}$ achieves $\epsilon$-DP \cite{dwork2006calibrating}.

To achieve $\epsilon$-DP, a simple mechanism is Laplace noise injection \cite{dwork2006calibrating}. We show the detailed noise injection mechanism in the following theorem.

\begin{theorem}\label{theorem2_}\rm({\textbf{Theorem 1 in \cite{dwork2006calibrating}}}) {We say a mechanism $\mathscr{B}$ achieves $\epsilon$-DP, if} $\mathscr{B}(D)$ satisfies
\begin{equation}
    \mathscr{B}(D)=q(D)+n,
\end{equation}
and $n$ is Laplace noise {with} probability density function $p(s)$:
\begin{equation}
\label{ps}
p(s)=\frac{1}{2\lambda}e^{-\frac{|s|}{\lambda}},
\end{equation}
in which $\lambda=\frac{\Delta f(D)}{\epsilon}$ {combines} the privacy parameter $\epsilon$ that describes the level of the privacy and the sensitivity $\Delta f(D)$ satisfying $\Delta f(D)\geq \max \|q(D_1)-q(D_2)\|$ for all neighbor datasets $D_1$ and $D_2$. 
\end{theorem}

\noindent {\textbf{Remark:} In our setting, the dataset $D$ can be viewed as the appliance states. The query function $q$ could be viewed as the meter data derived from the appliance states and appliances' mean energy consumption. The sensitivity $\Delta f$ is the meter data sensitivity defined in (\ref{dltf}) for $y_t$; and $\epsilon$ describes user's required level of the privacy. }

Thus, the Laplace noise injection mechanism helps us establish the connection between $\epsilon$-DP and NILM inference. 

Next, we construct a performance bound related to the privacy parameter $\epsilon$ and show how the level of privacy preservation {connects} to the performance {bound} of the NILM inference.

\subsection{Asymptotic Bounds for NILM with DP}
Recall that $\Delta f$ is the sensitivity for 
meter data $y_{t}$. After injecting the noise, instead of 
inferring $X_{t}$ using $y_{t}$, now we can only 
rely on the noisy data $y_{t}+n_{t}$, which yields optimization problem (P4).
\begin{equation}\label{The problem P}
\begin{aligned}
(\text{P4})\,\,\ \min\,\,&\|\Delta_{t}\|_{1}  \\
s.t.\,\,\,\,&K_{t}^\epsilon=|y_{t}+n_{t}-y_{t-1}-n_{t-1}|,\\
\,\,\,\,&\|\Delta_{t}P-K_{t}^\epsilon\|_{2}<\delta,\\
& \Delta_{t}\in [0,1]^{N}.
\end{aligned}
\end{equation}

Denote the solution to (P4) by $\Delta_{t}^\epsilon$, and 
the solution after rounding by $\bar{\Delta}_t^\epsilon$. 
Then, it is straightforward to apply Theorem 1 to (P4), and obtain the following proposition.

\vspace{0.1cm}
\begin{proposition}\label{proposition1}\rm
The difference between $\mathbf{E}[\bar{\Delta}_t^\epsilon]$ and the ground truth $\Delta_{t}^{0}$ is bounded. More precisely,
\begin{equation}\label{prop1}
    \|\mathbf{E}[\bar{\Delta}_t^\epsilon]-\Delta_{t}^{0}\|_{2}\leq (2+C(\mathbf{P}))\cdot\delta+|n_{t}-n_{t-1}|.
\end{equation}
Note that again $C(\mathbf{P})$ is {a} function of matrix $\mathbf{P}$, defined in Theorem 1. 
\end{proposition}

This proposition not only dictates that the error's upper bound increases almost linearly with the magnitude of noises, but also allows us to connect the DP parameters with the NILM inference accuracy.

Mathematically, we define inference accuracy $\alpha$ as follows:
\begin{equation}\label{acc_one}
\alpha = 1 - \frac{\mathbf{E}\|\bar{\Delta_{t}^\epsilon}-\Delta_{t}^{0}\|_{1}}{N}.
\end{equation}

To evaluate $\mathbf{E}[\alpha]$ over 
all possible injected noises $n_{t}$ and $n_{t-1}$, {we denote}
\begin{align}
    b = 2\left(N-(2+C(\mathbf{P}))\delta\right).
\end{align}

We can now establish the lower bound for NILM inference accuracy as Theorem 3 suggests.

\vspace{0.1cm}
\begin{theorem}\label{theorem3}\rm
The expectation of the inference accuracy $\alpha$ can be lower bounded. Specifically, 
\begin{equation}\label{theorem3_delta}
\begin{aligned}
\mathbf{E}[\alpha]\geq 1-\frac{4C(\mathbf{P})\delta\epsilon+8\delta\epsilon+3\delta}{4\epsilon N}
+\frac{A_{1}\epsilon+B_{1}}{4\epsilon N}e^{-\frac{2\epsilon b}{\delta}},
\end{aligned}
\end{equation}
where
\begin{align}
A_{1} = 2N-4\delta-2C(\mathbf{P})\delta,\\
B_{1} = 3\delta.
\end{align}
\end{theorem}

Note that a smaller $\epsilon$ implies a higher level of differential privacy.  In proving the lower bound of inference accuracy, the key is to identify that for
$\|\mathbf{E}[\bar{\Delta}_t^\epsilon]-\Delta_{t}^{0}\|_{2}$ in Proposition 1, there is a trivial upper bound, $N$. Hence, we can refine Proposition 1 to construct the lower bound for  $\mathbf{E}[\alpha]$.

As indicated by Theorem 3, the derived lower bound decreases at the rate of $\epsilon^{-1}$ when $\delta$ is small enough (i.e. the impact of the exponential term is neglectable).

To construct the upper bound, we need to exploit the structure of problem (P4). Specifically, standard mathematical manipulations of constraints in (P4) yield the following upper bound characterization:

\vspace{0.1cm}
\begin{theorem}\label{theorem4_}\rm
Denote $m = N\|\mathbf{P}\|_{2}+2\delta$ for notational simplification. The expected inference accuracy $\alpha$ can be upper bounded: 
\begin{equation}\label{theroem4_delta}
\begin{aligned}
\mathbf{E}[\alpha]\leq 1+\frac{A_{2}\epsilon^{2}+B_{2}\epsilon+C_{2}}{8\delta\epsilon N\|\mathbf{P}\|_{2}}e^{-\frac{2\epsilon m}{\delta}}-\frac{16\delta\epsilon^{2}+4\delta\epsilon+3\delta}{16\epsilon N\|\mathbf{P}\|_{2}}e^{-\epsilon},\nonumber
\end{aligned}
\end{equation}
where
\begin{align}
A_{2} = 4m^{2}-8\delta m-4mN\|\mathbf{P}\|_{2},\\
B_{2} = 6\delta m-8\delta^{2}-4\delta N\|\mathbf{P}\|_{2},\\
C_{2} = 3\delta^{2}.
\end{align}
\end{theorem}

Theorem 4 implies that the upper bound decreases at the rate of $o({\epsilon}^{-1}e^{-\epsilon})$. {That is, the upper bound would also decrease when $\delta$ becomes larger.} Hence, there is a certain gap between the lower bound and the upper bound. We want to emphasize that although the proof {(in Appendix \ref{Theorem 4})} is based on the compressive sensing framework, it can be generalized to many other NILM algorithms for the upper bound construction.

\section{Hierarchical Multi-shot NILM}
\label{muti}
In this section, we analyze the multi-shot NILM and propose our hierarchical decomposition algorithm. Specifically, we first introduce the multi-shot NILM formulation, identify its challenges and conduct the classical treatment. Then, we explain how to relax Assumption \ref{assumption3} with hierarchical decomposition. We conclude this section by analyzing the accuracy bounds for the corresponding {multi-shot} NILM inference.

\subsection{Multi-shot NILM Inference}
The special difficulty {in} multi-shot NILM inference, compared with one-shot NILM, is the temporal coupling {between inferences}. This issue, if not well addressed, could lead to large cumulative errors, and ultimately lead to the cascading inference failure\footnote{{See the cascading effect over network in \cite{easley2012networks} for more detailed discussion.}}. Therefore, in this part, after the problem formulation, we intend to {design an effective algorithm to contain this error.}

The multi-shot NILM inference problem is a straightforward generalization of the one-shot NILM inference. It seeks to infer the appliance switch events over a period of $T$ from the meter data sequence $\boldsymbol{Y}=[y_1,..y_T]$. The inference also relies on the {vector of} mean energy consumption $\mathbf{P}$, defined in Eq. (\ref{eq1}). We denote $\boldsymbol{X}=[X_1,...,X_T]$ the state probability matrix, where each $X_t$ captures the probability of appliances states at time $t$. To establish the asymptotic bounds for our multi-shot NILM inference accuracy, we denote $\boldsymbol{X}^0=[X^0_1,...,X^0_T]$ the ground truth of switching events over $T$, and $\bar{\boldsymbol{X}^{\epsilon}}=[\bar{X_1^{\epsilon}},..,\bar{X_T^{\epsilon}}]$ the inference results returned by our proposed algorithm.

It is worth noting that the situation becomes even more {complicated} if we conduct rounding at each time slot. We submit that it is valuable to keep the inference results obtained in solving problem (P3) without rounding as these results can be {interpreted} as the probability of switching. This allows us to derive the state probability matrix $\boldsymbol{X}=[X_1,...,X_T]$.
Note that, after obtaining the matrix $\boldsymbol{X}$, we can no longer conduct the rounding naively: there is more information about the actual meter readings. We could utilize the simple error correcting methods to decipher the embedded information.

Specifically, the error correcting aims to make the binary adjustments to reduce the approximation error according to the actual meter data. Though it could be possible for us to consider more complex methods for error correcting rather than direct adjustment, they make little difference in terms of the worst case bound analysis.

We describe the classical treatment to multi-shot NILM inference in Algorithm \ref{mutishotcs}, where we adopt the notion of the Hadamard Product, which is defined for two matrices $\boldsymbol{A}$ and $\boldsymbol{B}$ as follows:

\vspace{0.1cm}
\begin{definition}\label{def2}\rm
For two matrices $A$ and $B$ of the same dimension $m\times n$, the Hadamard product $A\bigodot B$ is the $m\times n$ matrix $C$ with elements given by
\begin{align}
    c_{ij}=a_{ij}b_{ij},\ \forall 1\le i\le m,\ \forall 1\le j\le n.
\end{align} 
\end{definition}

There is a simple way to construct {rough} inference accuracy bounds directly from Theorem 3 and Theorem 4. The trick is to examine the {rounding procedure in Algorithm \ref{mutishotcs}} and to connect it with the bounds for one-shot NILM inference. Formally, we define the accuracy of the classical algorithm for multi-shot NILM inference by $\alpha_m$:
\begin{equation}\label{acc}
{\alpha}_m = 1 - \frac{\|\bar{\boldsymbol{X}^{\epsilon}}-\boldsymbol{X}^{0}\|_{1}}{NT}.
\end{equation}

Denote the lower bound and upper bound for the one-shot NILM inference accuracy by $b(\delta,\epsilon)$ and $B(\delta,\epsilon)$, respectively {(see Theorem \ref{theorem3} and \ref{theorem4_})}.
Recall that $\epsilon$ is the DP's parameters {(see Theorem \ref{theorem2_})} and $\delta$ describes the sensitivity of the meter data for each period $y_t$ {(see Eq. (\ref{dltf}))}. 

With these definitions, we can characterize the lower and upper bounds for $\mathbf{E}[\alpha_m]$:

\vspace{0.1cm}
\begin{theorem}\label{theorem5_}\rm
The expected inference accuracy $\mathbf{E}[\alpha_m]$ for the multi-shot NILM problem can be lower and upper bounded. Specifically, we have
\begin{equation}
1-\frac{(T-1)(1-b(\delta,\epsilon)N)}{2}\leq\mathbf{E}[\alpha_m]\leq 1-\frac{1-B(\delta,\epsilon)}{T},
\end{equation}
where $b(\delta,\epsilon)$ and $B(\delta,\epsilon)$ are the lower bound and upper bound for the one-shot NILM inference accuracy. 
\end{theorem}

\renewcommand{\algorithmicrequire}{\textbf{Input:}}  
\renewcommand{\algorithmicensure}{\textbf{Output:}} 
\begin{algorithm}[!ht]
\caption{Multi-shot Inference}
\label{mutishotcs}
\begin{algorithmic}[1]
\REQUIRE The initial state $X_{0}$;\\
The meter reading sequence $\boldsymbol{Y}=y_{0},y_{1},...,y_{t}$;\\
The appliances' number $N$;\\
The appliances' mean power vector $\mathbf{P}$;

\ENSURE The appliance state sequences $\bar{X^\epsilon_{1}},...,\bar{X^\epsilon_{T}}$;
\STATE Get the differences between time points of the meter readings as $k_{1},...,k_{t}$, $K_t = y_t - y_{t-1}, t = 1,...,T;$
\STATE Solve the optimization problem (\ref{The problem P3}) for each $t$ and get the approximation solution
$\Delta_{1},...,\Delta_{T}$;
\FOR{time $t$ from $1$ to $T$}
\STATE $X_{t} = X_{t-1} \bigodot (1-\Delta_{t-1})+(1-X_{t-1}) \bigodot \Delta_{t-1}$;
\ENDFOR
\FOR{time $t$ from $1$ to $T$}
\STATE Do rounding on $X_t$ and derive the vector $\bar{X_t^\epsilon}$
\IF{$\bar{X_t^\epsilon}\mathbf{P}>y_{t}$}
\STATE $k=0$;
\WHILE{$\bar{X_t^\epsilon}\mathbf{P}>y_{t}$ and $k<N$}
\STATE find $k^{th}$ largest appliance $j$ and set 
$\bar{X_{t}^{\epsilon j}}=0$;
\STATE$k=k+1$;
\ENDWHILE
\ELSIF{$\bar{X_t^\epsilon}\mathbf{P}<y_{t}$}
\STATE $k=0$;
\WHILE{$\bar{X_t^\epsilon}\mathbf{P}<y_{t}$ and $k<N$}
\STATE find $k^{th}$ smallest appliance $j$ and set
$\bar{X_{t}^{\epsilon j}}=1$;
\STATE$k=k+1$;
\ENDWHILE
\ENDIF
\ENDFOR
\RETURN $\bar{X^\epsilon_{1}},...,\bar{X^\epsilon_{T}}$.
\end{algorithmic}
\end{algorithm}

\subsection{Hierarchical Decomposition}\label{hmutialgo}

Albeit that we have deployed the error correction method, the two bounds in Theorem $5$ are still rather loose. For example, as $T$ increases, the lower bound could quickly approach $0$ while the upper bound would quickly approach $1$. This is because we still need to take into account the cascading effects in the union bound though we have {made} some efforts to {contain the error}. Another reason is due to Assumption \ref{assumption3}. We may relax this assumption for improved bounds.

\par The idea is simple and straightforward. Intuitively, to relax Assumption 3, we need to conduct a hierarchical decomposition for different proper appliance sets. To {obtain the proper} sets, we first sort all the appliances in $\mathcal{A}$ according to their mean energy consumption {levels}. 
Then, starting with the least energy-consuming appliance, we could form a number of proper sets that satisfy Assumption 3.

To construct the sets, we use the greedy policy. For a certain set with $S$ appliances, the condition that makes the $(S+1)^{th}$ appliance join the current set instead of creating a new set is as follows: 
\begin{align}
\sum\nolimits_{i=1}^{\lfloor\frac{S}{2}\rfloor+1}P_{i}-2\delta\geq\sum\nolimits_{i=1}^{\lfloor\frac{S}{2}\rfloor-1}P_{S+1-i}+P_{S+1}, \label{cD2}
\end{align}
where the subscript $i$ in $P_i$ also illustrates its ascending order in the set.
\vspace{0.1cm}
\begin{proposition}\label{proposition2}\rm
The set generation criteria (\ref{cD2}) guarantees the validity of Assumption \ref{assumption3} for each proper generated set.
\end{proposition}
\vspace{0.1cm}

We call each set a hierarchy, which establishes the basis for our hierarchical multi-shot NILM inference. We apply Algorithm 1 to each hierarchy in the descending order. We characterize the whole procedure in Algorithm \ref{hmutishot-cs}. 

\par We submit that the hierarchical decomposition helps refine asymptotic bounds for $\mathbf{E}[\alpha_m]$ in that it characterizes the connection between hierarchies. This allows us to derive a recursive formulation to estimate the overall inference accuracy. Define inference accuracy for hierarchy $i$ as $\alpha_i$:

\begin{equation}\label{acc_hier}
\alpha_i = 1 - \frac{\|\bar{\boldsymbol{X}^{\epsilon}[i]}-\boldsymbol{X}[i]^{0}\|_{1}}{N_{i}T},
\end{equation}
where $N_{i}$ denotes the number of appliances in hierarchy $i$; $\boldsymbol{X}^{\epsilon}[i]$ and $\boldsymbol{X}[i]^{0}$ denote the inferred result and the ground truth of the appliances in hierarchy $i$, respectively. 
We also define the {mean} power matrix for hierarchy $i$ as $\mathbf{P}_{i}$ and the minimum power in hierarchy $i$ as $P^{i}_m$.
For simplicity, we also define the maximal summation of $U$ appliances whose power is smaller than $P^{i}_m$ as $P^{i}_U$, in which $U$ is the sparsity of the switching events as we defined in Assumption 1.

Now we can formally introduce our bounds characterization theorem for hierarchical multi-shot NILM inference, with the help of Theorem 5. For simplicity, define
\begin{align}\label{bound_use}
    &b_m(\delta,\epsilon) = 1- \frac{(T-1)(1-b(\delta,\epsilon)N)}{2},
    \\&B_M(\delta,\epsilon) = 1- \frac{1-B(\delta,\epsilon)}{T}.
\end{align}
Then, we can {show} that

\vspace{0.1cm}
\begin{theorem}\label{theorem6_}\rm
The expected inference accuracy for each hierarchy $i$, $\mathbf{E}[\alpha_i]$, can be lower bounded by $m_i$ and upper bounded by $M_i$.
Specifically,
\begin{equation}
\begin{aligned}
    m_i=b_m\left( \delta+\frac{2}{2+C(\mathbf{P})}\delta_i',\epsilon \right),\\ M_i=B_M(\delta+\delta_i',\epsilon),
\end{aligned}
\end{equation}
where
\begin{equation}
\delta_i'=\frac{P_{U}^i}{2}+\sum_{k=1}^{i-1}N_{k}T(1-m_{k})\|\mathbf{P}_k\|_2.
\end{equation}
Note that, by definition of summation, $\delta_1' = \frac{P_{U}^1}{2}$.
\end{theorem}

\renewcommand{\algorithmicrequire}{\textbf{Input:}}  
\renewcommand{\algorithmicensure}{\textbf{Output:}} 
\begin{algorithm}[t]
\caption{Hierarchical Multi-shot Inference}
\label{hmutishot-cs}
\begin{algorithmic}[1]
\REQUIRE The initial state $X_{0}$;\\
The meter reading sequence $y_{0},y_{1},...,y_{T}$;\\ 
The number of appliances $N$;\\ 
The appliances power vector $\mathbf{P}$;
\ENSURE The appliances states sequences $\bar{\boldsymbol{X^\epsilon}} = \bar{X^\epsilon_{1}},...,\bar{X^\epsilon_{T}}$;
\STATE \textbf{Decomposition:}
\STATE Rank the power vector $\mathbf{P}$ ascendingly: $P_{1},...,P_{N}$;
\STATE $k=1$, $C=\varnothing$, $S=\varnothing$; 
\FOR{$i$ from 1 to $N$}
\IF{$|C|=0$ or $1$}
\STATE $C\leftarrow{C\cup\{P_i\}}$;
\ELSE
\STATE For the items in $C$, rank them ascendingly: $c_{1},...,c_{|C|}$;
\IF{ $\sum_{j=1}^{\lfloor\frac{|C|}{2}\rfloor+1}c_{j}-2\delta\geq\sum_{j=1}^{\lfloor\frac{|C|}{2}\rfloor-1}c_{|C|+1-j}+P_{i}$}
\STATE $C\leftarrow{C\cup\{P_i\}}$;
\ELSE
\STATE $S\leftarrow{S\cup\{C\}}$;
\STATE $C = {P_i}$;
\ENDIF
\ENDIF
\ENDFOR 
\STATE \textbf{Decoding:}
\STATE Rank $S$ descendingly with respect to the largest element in each set;
\FOR{$S_{i}$ in $S$}
\STATE Construct the appliances in $S_{i}$'s power vector $P_{S_{i}}$;
\STATE $X_0[S_i]$ to denote the initial states of the appliances in the set $S$; 
\STATE Conduct Multi-shot Compressive Sensing with $X_{0}[S_i]$, $\boldsymbol{Y}$, $|S_{i}|$, $P_{S_{i}}$ and derive $\bar{\boldsymbol{X}^\epsilon}[S_i]$
\STATE $\boldsymbol{Y} = \boldsymbol{Y}-\bar{\boldsymbol{X}^\epsilon}[S_{i}]P_{S_{i}}$;
\ENDFOR
\RETURN $\bar{\boldsymbol{X}^\epsilon}=\bar{X_{1}^\epsilon},...,\bar{X_{T}^\epsilon}$.
\end{algorithmic}
\end{algorithm}

Specifically, suppose we have $K$ hierarchies in the problem, the overall accuracy $\alpha_H$ can be {recursively} obtained as follows:
\begin{equation}\label{acc_total}
\mathbf{E}[\alpha_H]=\frac{\sum_{i=1}^{K}\mathbf{E}[\alpha_i]N_i}{\sum_{i=1}^{K}N_i}.
\end{equation}
The associated bounds for our multi-shot compressive sensing algorithm with hierarchical decomposition could also be derived straightforwardly with the help of Theorem 6.

\vspace{0.1cm}
\noindent {\textbf{Remark}: It is possible to generalize our framework to the non-binary multiple discrete states case. The key modification is to characterize the state of each appliance with a vector of random variables, instead of a single random variable. However, in this paper, we choose to focus on a stylized model, which only considers the binary state. This stylized model, by simplifying the mathematical details, helps to sharpen our understanding of the physical insights into the DP parameters in the electricity sector.}

\section{Numerical Studies}\label{experiments}
In this section, we design a sequence of numerical studies to highlight the effectiveness of our proposed bound characterization.
The effectiveness lies in the consistent magnitude with the empirical trends of the inference accuracy (in DP parameter $\epsilon$). 

We conclude this section by illustrating that our derived asymptotic bounds are not only effective to the compressing sensing framework, but also effective to {characterize the performance of} many other NILM algorithms.

\subsection{Overview of the Datasets}

{In this paper, we use three publicly available datasets.}
\begin{itemize}
    \item {The UK-DALE dataset \cite{kelly2015the} records the electricity consumption from five households in UK, with a sampling rate of 6 Hz for each individual appliances. For performance evaluations, we choose the data segment from 2013/05/26 21:03:14 to 2013/05/27 0:54:50. }
    
    \item {The TEALD dataset \cite{makonin2018rae:} is collected by Rainforest EMU2 in Canada. With a sampling rate of 1 Hz, we use the data segment from 2016/02/09 16:00:00 to 2016/02/10 15:59:59 for performance evaluation.}
    
    \item {The Redd dataset \cite{Kolter2011REDDA} is also commonly used for NILM performance evaluation. We use the data from 2011/04/18 21:22:13 to 2011/05/25 03:56:34 with a sampling rate of 3 HZ.}
\end{itemize}

\subsection{One-shot NILM Inference}

We use the dataset of building 4 in UK-DALE 
dataset \cite{kelly2015the} to conduct the first numerical 
analysis. In this building, there are 8 appliances. 

\begin{figure}[!t]
\centering
\includegraphics[width=2.7 in]{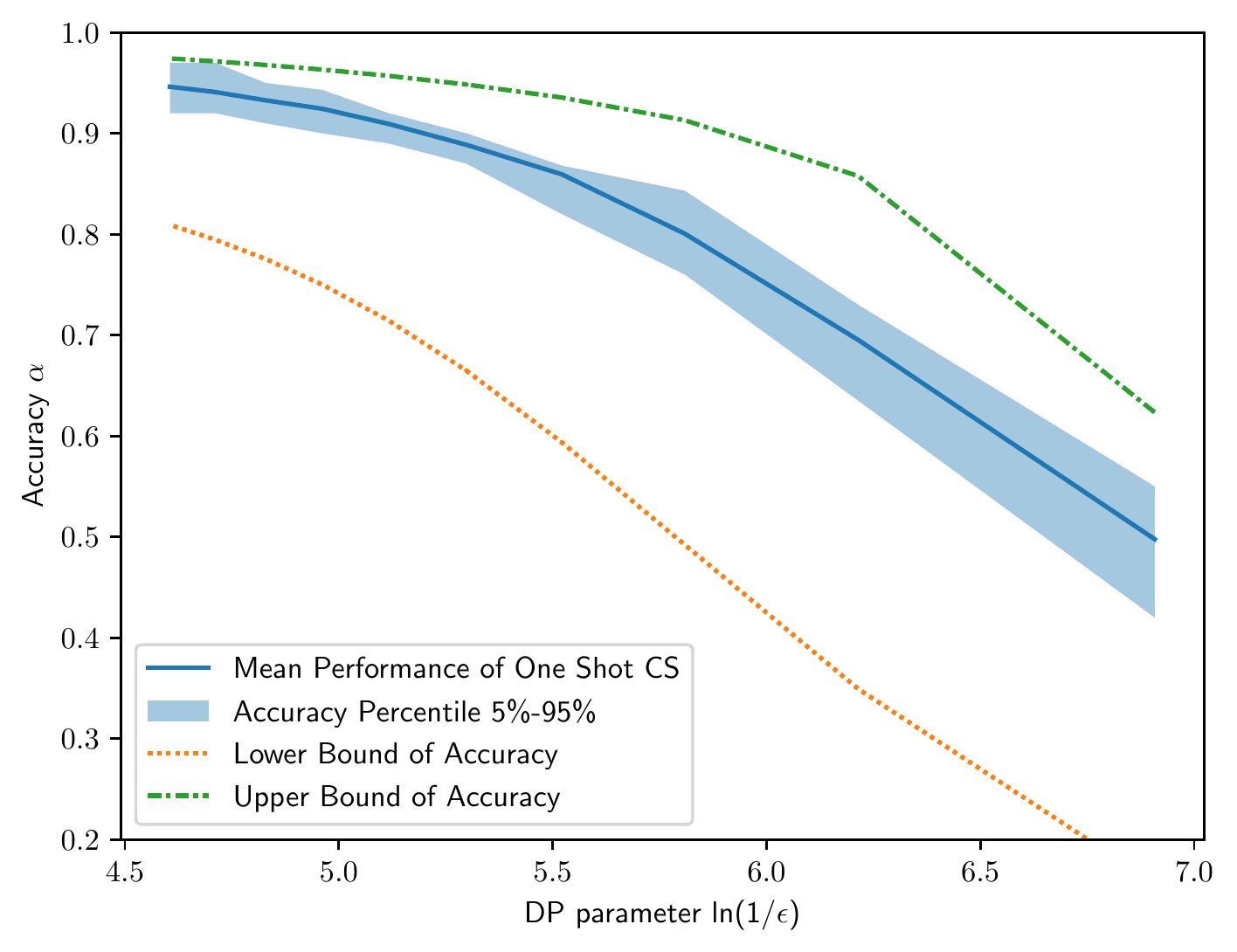}\vspace{-0.2cm}
\caption{One-shot NILM Inference Accuracy in $\ln(\frac{1}{\epsilon})$.\vspace{-0.5cm}}
\label{bound}
\end{figure}

Using the one-shot NILM inference based on 
compressive sensing, we evaluate the inference 
performance with increasing level of DP. 
Fig. \ref{bound} plots the theoretical 
lower bound, the actual performance traces, 
and the theoretical upper bound to evaluate our theorems. The lower bound is not tight 
mainly due to the sensitivity and parameter 
$C(\mathbf{P})$. 
The sensitivity is not tight in that it is time-varying. In some time slots, some appliances are simply idle. However, they all contribute to the sensitivity.
The parameter $C(\boldsymbol{P})$ is not tight due to exactly the same reason.
The meter data suggests $\delta$ be $2$, and $C(\boldsymbol{P})$ should be $0.015$. 
Nevertheless, our lower bound is tight in terms of magnitude. The mean of the actual inference accuracy indeed decreases at the rate of $\epsilon^{-1}$.

On the other hand, our proposed upper bound seems tight for this case study. We want to emphasize that this upper bound may not always be tight in that we assume the existence of an oracle in the upper bound construction.
Hence, this upper bound would be more ideal to serve as a benchmark for all possible NILM algorithms. Hence, from Fig. \ref{bound}, we can conclude that compressive sensing is a good approach for one-shot NILM inference.

\subsection{Multi-shot NILM Inference}\label{multi_sec}
\begin{figure}[!t]
  \centering
  \includegraphics[width=2.7 in]{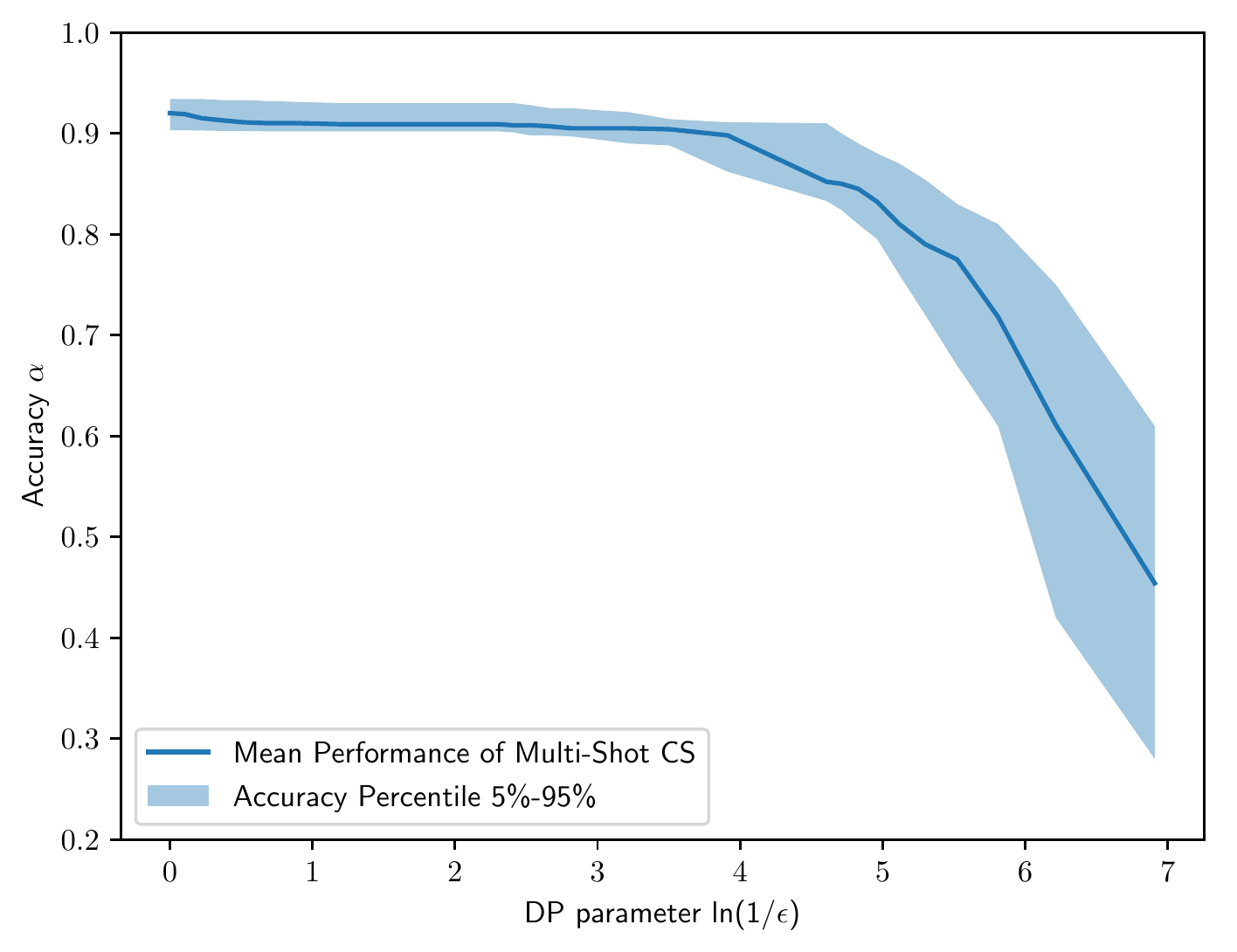}\vspace{-0.2cm}
  \caption{Multi-shot NILM Inference Accuracy in $\ln(\frac{1}{\epsilon})$.}
  \label{mutishot}
  \end{figure}

We then use the UK-DALE dataset to validate our Algorithm \ref{mutishotcs} which relies on Assumption 3. The inference accuracy is shown in Fig. \ref{mutishot}. In fact, our proposed approach performs {remarkably good}, and the inference accuracy trend shows a similar decreasing rate as Theorem 6 dictates. We want to emphasize that, all of our figures are plotted on a half log scale, which is ideal to verify the decreasing rate of $\epsilon^{-1}$.

\begin{figure}[!t] 
  \centering
  \includegraphics[width=2.7 in]{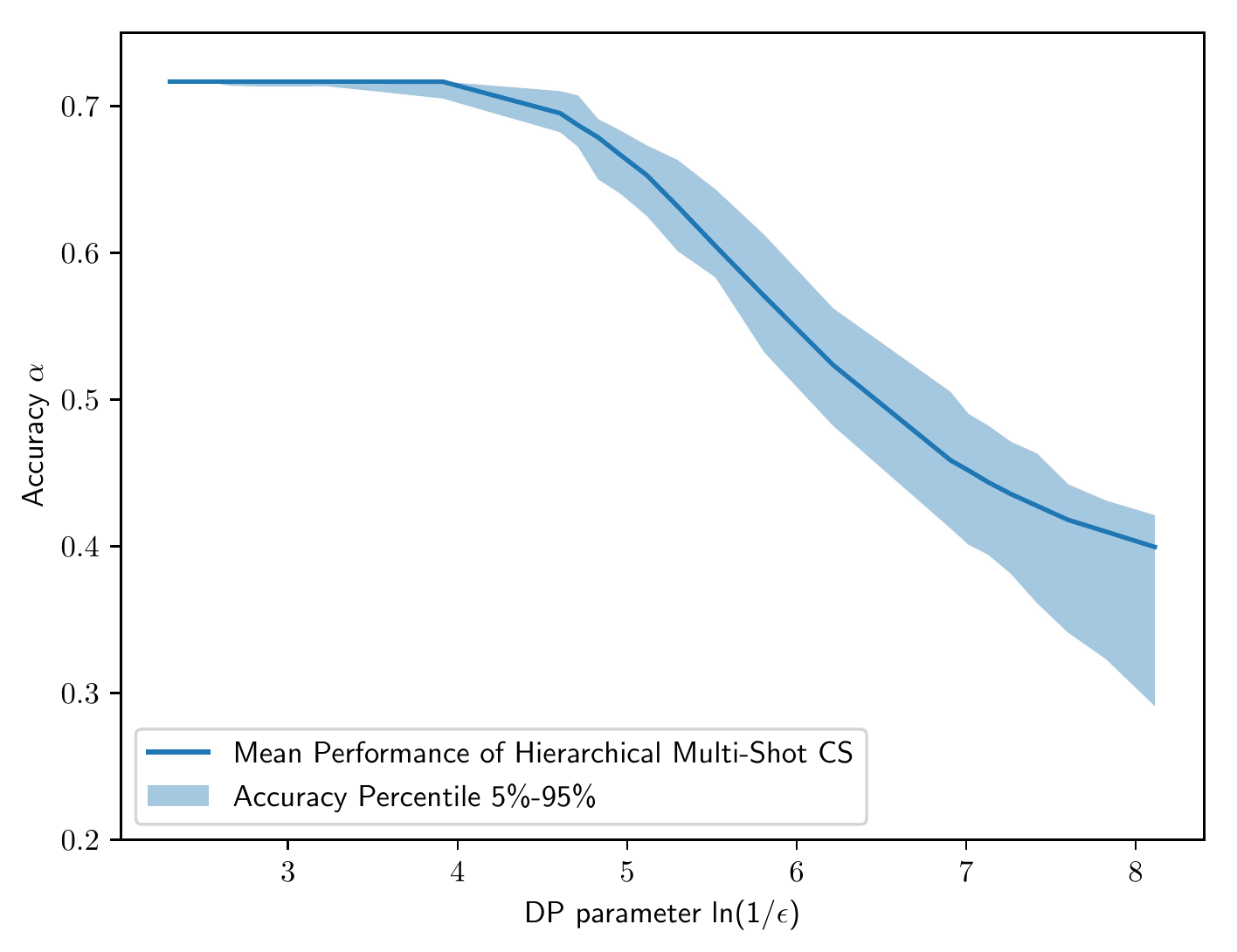}\vspace{-0.2cm}
  \caption{Hierarchical Multi-shot NILM Inference Accuracy in $\ln(\frac{1}{\epsilon})$}
  \label{hier}
  \end{figure}

We use the TEALD dataset to further demonstrate the effectiveness of our hierarchical decomposition.
There are 13 appliances in the dataset, with diverse energy consumptions.
We group them into four hierarchies and conduct the NILM inference. Fig. \ref{hier} plots its inference accuracy trend in the DP parameter, which again highlights the remarkable performance of our proposed algorithm, and verifies the effectiveness of our derived bounds. 
During the simulation, we observe that our hierarchical algorithm seems to be quite robust against noises, which may lead to interesting theoretical results, though a detailed discussion is beyond the scope of our work.

\subsection{Robustness of Noise Injection Methods}
{We examine the robustness of our conclusion through numerical studies. Specifically, to understand if our conclusion only holds for Laplace noise injection. We consider other noise injection methods, which also achieves $\epsilon$-DP. Here, we employ the staircase mechanism, described in Algorithm \ref{staircase}, which achieves $\epsilon$-DP \cite{geng2014optimal}.}

\renewcommand{\algorithmicrequire}{\textbf{Input:}}  
\renewcommand{\algorithmicensure}{\textbf{Output:}} 
\begin{algorithm}[t]
\caption{Generation of Staircase noise \cite{geng2014optimal}}
\label{staircase}
\begin{algorithmic}[1]
\REQUIRE $\epsilon$, $\Delta f$;
\ENSURE the staircase noise $n$;
\STATE $\gamma \leftarrow 1/(1+e^{\epsilon/2})$;
\STATE Generate a discrete random variable $S$ $\rm{Pr}(S=1)=\rm{Pr}(S=-1)=\frac{1}{2}$;\\
\STATE Generate a geometric random  variable $G$ with $\rm{Pr}(G=i)=(1-b)b^i$ for integer $i \geq 0$, \\
where $b=e^{-\epsilon}$;
\STATE Generate a random variable $U$ from a uniform distribution in $[0,1]$;
\STATE Generate a binary random variable $B$ with $\rm{Pr}(B=0)=\frac{\gamma}{\gamma+(1-\gamma)b}$ and $\rm{Pr}(B=1)=\frac{(1-\gamma)b}{\gamma+(1-\gamma)b}$;
\STATE {\small{$n\!\leftarrow\! S((1\!-\!B)((G\!+\!\gamma U)\Delta f)\!+\!B((G\!+\!\gamma\!+\!(1\!-\!\gamma)U)\Delta f))$}};
\RETURN $n$.
\end{algorithmic}
\end{algorithm}

\begin{figure}[!t]
  \centering
  \includegraphics[width=2.7 in]{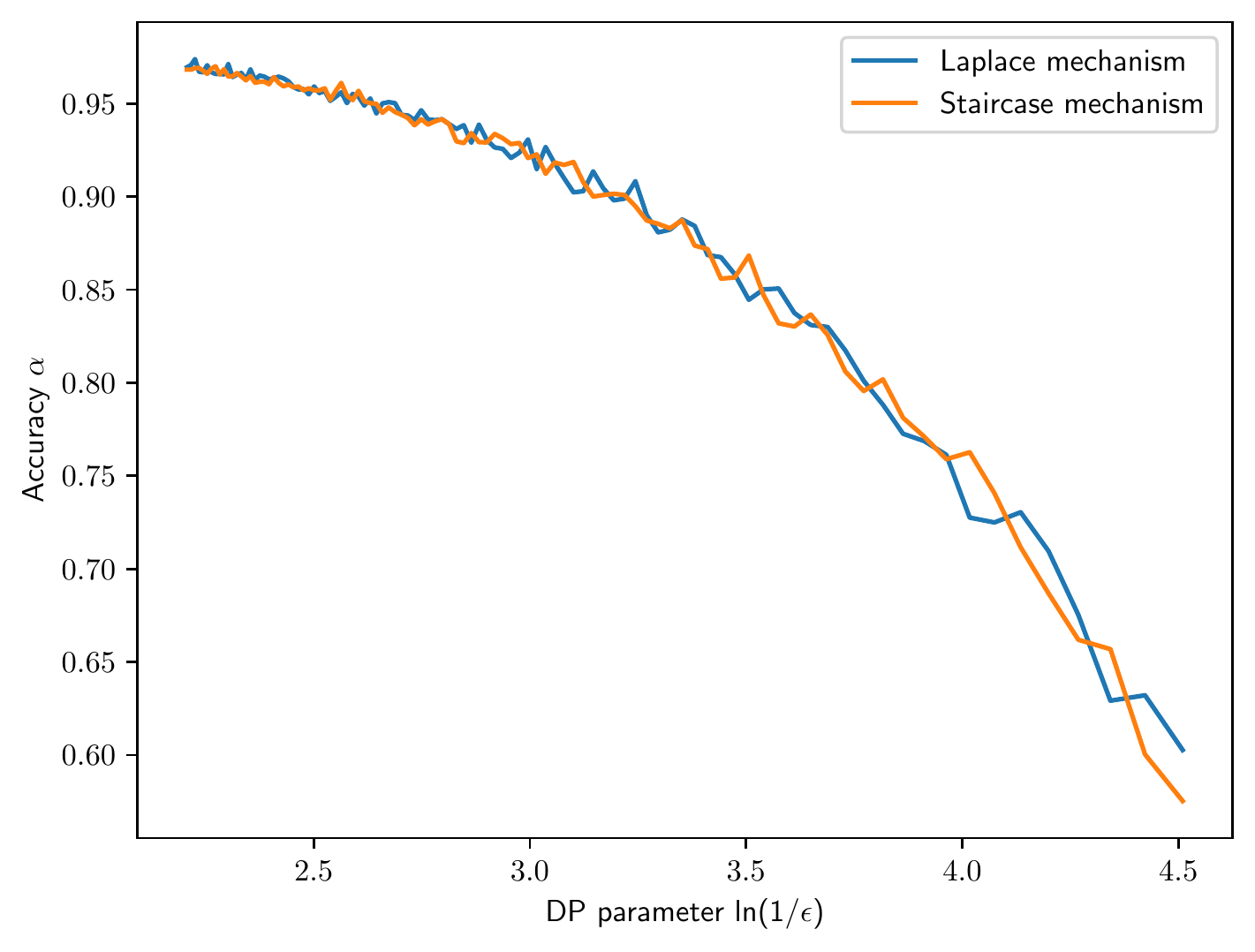}\vspace{-0.2cm}
  \caption{Robustness Analysis}
  \label{noisetype}
  \end{figure}

{We follow the same routine in Section \ref{multi_sec} to conduct the numerical study and compare the results in Fig. \ref{noisetype}. It is clear that different noise injection methods lead to similar relationships between the DP parameters and the inference accuracy. This further illustrates the robustness of our conclusion against noise injection methods.}

\subsection{Beyond Compressive Sensing}

We can now safely conclude that, in the compressive sensing framework, the inference accuracy heavily relies on the DP parameter $\epsilon$.
Specifically, we observe the inference accuracy decreases linearly with $\ln{(\frac{1}{\epsilon})}$ as our figure shows when $\ln{(\frac{1}{\epsilon})}$ is large. We now try to generalize this conclusion to other common frameworks in NILM literature. Specifically, we conduct numerical studies to empirically {compare how} DP's parameter {affects} the accuracy of common NILM algorithms,
including sparse Viterbi algorithm for super-state HMM (Sparse-HMM), aided linear Integer Programming (ALIP), linear Integer Programming (IP), Recurrent Neural Network (RNN), CO, FHMM, and our proposed algorithm (Multi-shotCS).

We conduct numerical comparison on Building 1 in the Redd dataset.
For supervised learning models, we divide the dataset for training (before 2011-04-03) and for validation (after 2011-04-03).
Then, we inject noises to the {validation} set for performance evaluation.
For unsupervised model (including our approach), we {directly} inject noises to the meter data.

{
We explain the details about the hyperparameter settings as follows.
In fact, most of the settings follow the seminal works.}
\begin{itemize}
    \item {When implementing the SparseHMM algorithm, we follow the hyperparameters in the seminal Redd dataset numerical study in \cite{makonin2016exploiting}. Specifically, the number of the maximal super state is set to be $4$; and the state number choice parameter $\epsilon$ is set to be $0.00021$.}
    \item {As for ALIP and IP algorithms, we follow all the parameters in \cite{bhotto2017load}, including states, specific ratings and the state transient values. We choose the zero initialization for these two algorithms.}
    \item {For RNN implementation, we follow the RNN structure in \cite{kelly2015neural}. Specifically, we first adopt the 1D convolution layer to handle the input data. And then use two bidirection LSTM model and a fully connected layer to derive the output. We adopt the Adam optimizer and the initial learning rate of $0.01$ for training. The number of training epochs is set to be $5$ and the batch size is set to be $128$.}
    \item {For CO and FHMM, we again adopt the classical settings in NILMTK \cite{batra2014nilmtk}. The detailed settings are also provided in the online document \cite{nilmtk}.}
    \item {For our proposed Multi-shot compressive sensing algorithm, we evaluate the meter data and suggest $\delta$ should be $20$. This is derived by observing the maximal fluctuations in the historical meter data during the period without any switch event.}
    \end{itemize}

\begin{figure}[!t]
\centering
\includegraphics[width=2.7 in]{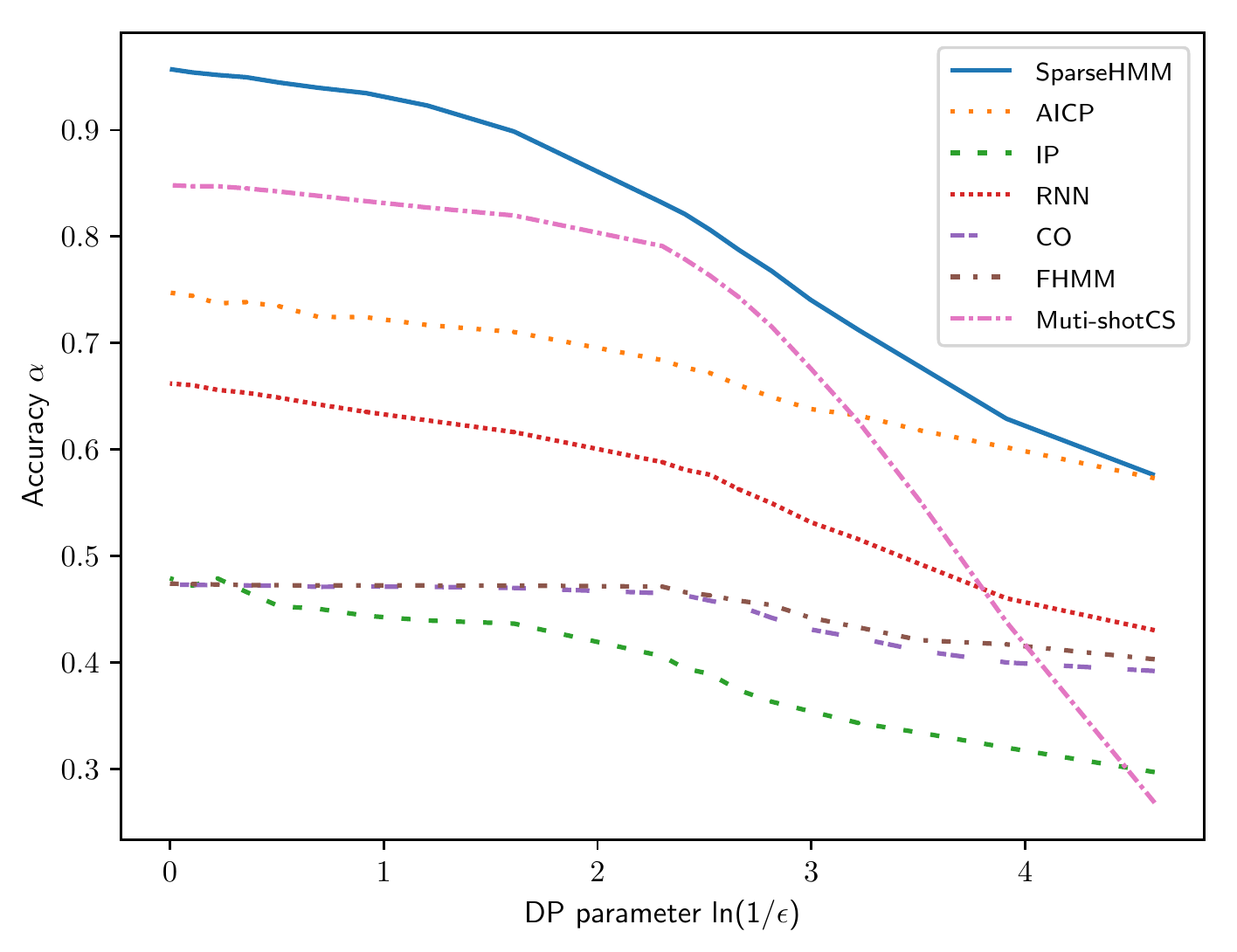}\vspace{-0.2cm}
\caption{NILM inference accuracy comparison among $7$ algorithms.\vspace{-0.5cm}}
\label{otheralgorithm}
\end{figure}

{As indicated by Fig. \ref{otheralgorithm}, among all the algorithms, SparseHMM achieves the best performance as it makes full use of the sparsity assumption and exploits the temporal correlations through the hidden Markov chain. They are also the underlying guarantee for the remarkable performance of our proposed framework. However, the utilization of temporal correlation makes both algorithms vulnerable to the noise injection due to the cascading inference failure. While ALIP outperforms the conventional IP approach, both of them show strong robustness to noise injection. This is because they do not rely on the sparsity assumption, and the median filter process in the two algorithms contributes to smooth out the fluctuations incurred by noise injection. The ordinary performance of RNN is due to the limited training data and the overfitting issue. The fewer assumption utilized by classical methods (CO and FHMM) makes them more robust to noise injection at the cost of low inference accuracy.}

The key observation is that the inference accuracy of each of the seven algorithms exhibits piecewise linear relationship with $\ln(\frac{1}{\epsilon})$, though with different breaking points and slopes. The slopes can serve as the indicators to evaluate each algorithm's robustness to Laplace noise (and hence its difficulty to achieve privacy preserving).

Nonetheless, our proposed bounds provide the correct magnitude to explain the physical meaning of DP parameters in most common frameworks for NILM, which could be inferred from Fig. \ref{otheralgorithm} empirically. Therefore, in the practical utilization of DP, our theoretical performance bounds could offer magnitude estimation for the effects of the different DP levels. 
Moreover, these empirical results also inspire us to derive a performance bound for these NILM algorithms for further discussion.

\section{Conclusion}\label{conclusion}
In this paper, we seek to offer the 
explicit physical meaning of $\epsilon$ 
in $\epsilon$-DP. 
{We firstly theoretically derive how the level of DP affects the performance guarantee for NILM inference. Then, we use extensive numerical studies to highlight that our theoretical results are effective not only for in the compressive sensing framework, but also for many other algorithms. These results could be helpful for the potential new business models for privacy preserving in the electricity sector.} 

This work can be extended in many ways. For example, 
it will be interesting to derive a tighter upper bound, or
a tighter lower bound for the inference accuracy. 
Then a more detailed analysis for the behavior individual appliance under DP framework could be conducted.
It is also an interesting topic to analyze the robustness of different
NILM algorithms when the privacy preservation is implemented. 

\bibliographystyle{unsrt}
\bibliography{ref}

\appendix

\subsection{DP May Affect Behavior Analysis}\label{inflence_and_implement}

{
It is believed that the load profile is an important pattern to characterize consumer behavior. Yu \emph{et al.} further show that the load profile can uniquely determine the system's marginal cost in serving each kind of consumer \cite{yu2017good}, and suggest that the k-means clustering based on $l_1$-norm is the suitable measure to cluster the users with similar system's marginal serving cost.}

\begin{figure}[!t]
  \centering
   \includegraphics[width=2.7 in]{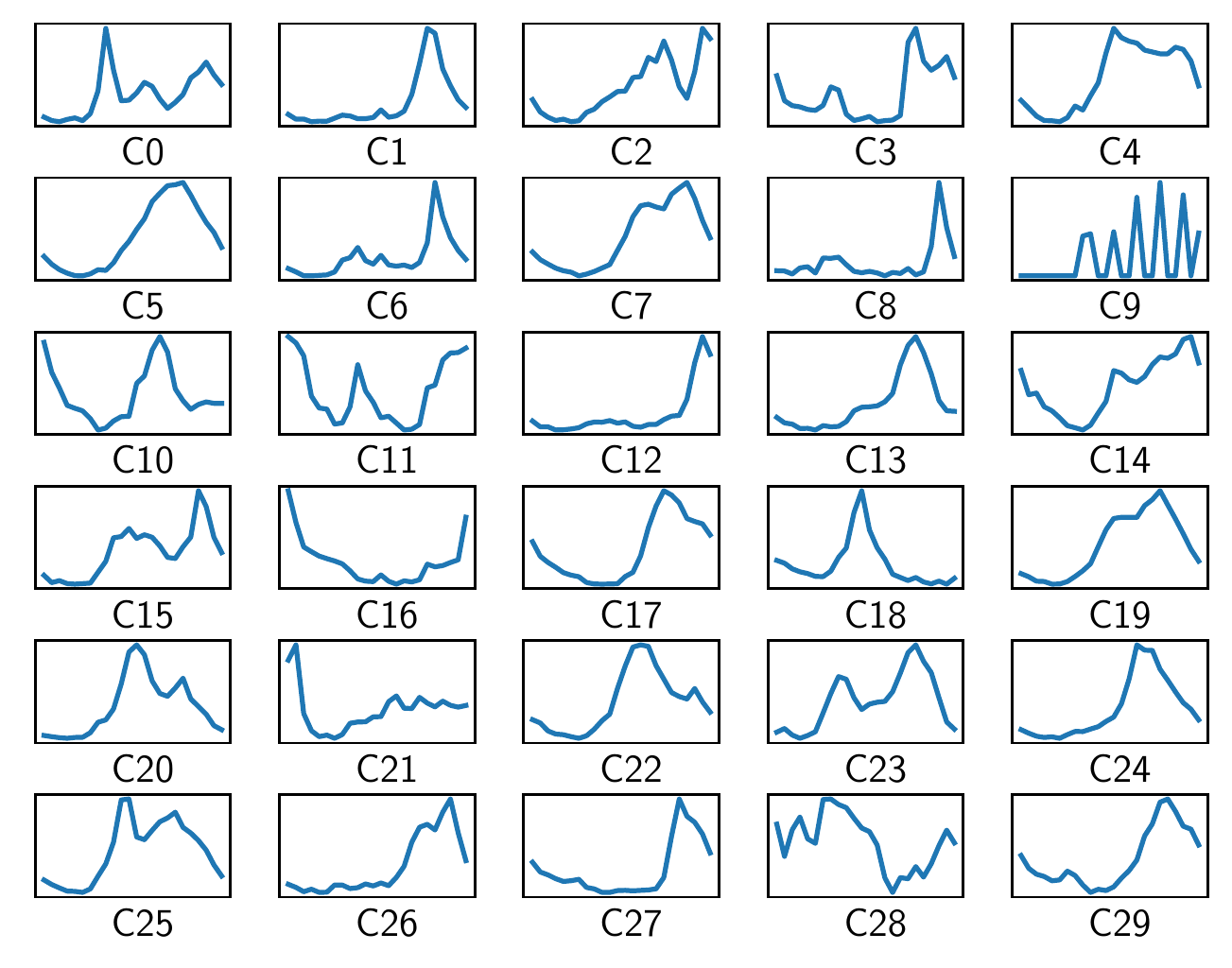}\vspace{-0.2cm}
  \caption{The profile of the clustering center}
  \label{clustering2}
  \end{figure}

{
To empirically highlight how the DP parameters may blur the behavior analysis, we use the Pecan Street dataset \cite{DemandData}, and randomly select $40$ users' $3$-month daily energy consumption profiles (from May 1 to Aug. 8, 2015). For a larger dataset, we combine all the profiles together and directly conduct the behavior analysis on this combined dataset. We select the number of clusters to be $30$. The initial clustering results are indicated in Fig. \ref{clustering2}.
}

\begin{figure}[htbp]
  \centering
  \includegraphics[width=3.3 in]{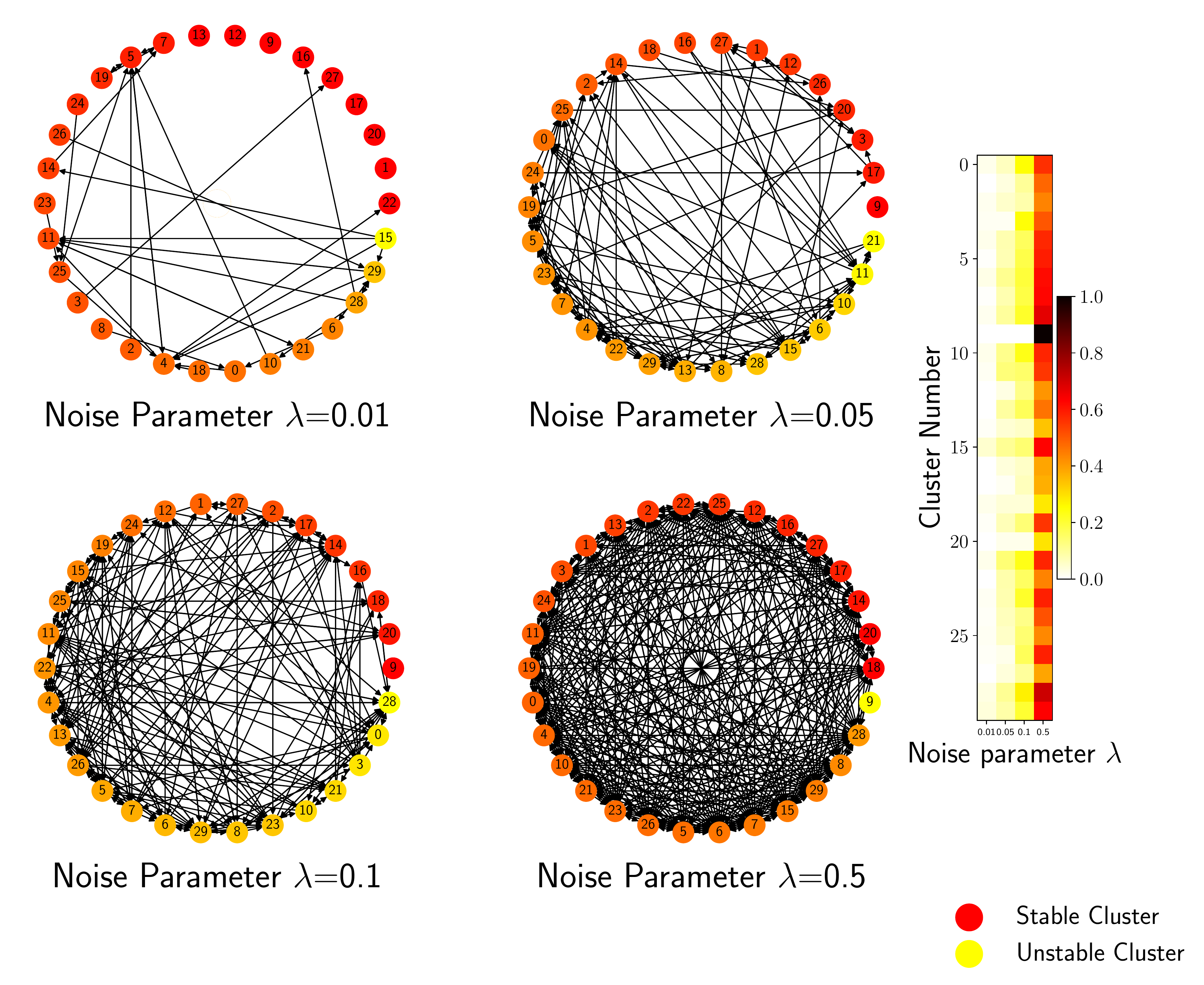}\vspace{-0.2cm}
  \caption{The clusters interchanges influenced by different levels of noise injection}
  \label{clusterchanges}
  \end{figure}

{
We inject the Laplace noise into every single consumer's load profile. We set the DP-parameter $\lambda$ to be $0.01, 0.05, 0.1, 0.5$, respectively to examine how this parameter may change the clustering result. Fig. \ref{clusterchanges} plots such impacts. Specifically, if a cluster has little users deviating to other clusters after noise injection, we call it a stable cluster. The color of each cluster represents such stability. The arrows in the figure show the trajectories of cluster interchanges. Obviously, higher privacy requirement (higher $\lambda$) leads to higher probability that users deviate from the original cluster.}

{    
To better connect the DP parameter with accuracy, we can estimate the probability that each consumer may stay in the original cluster and derive the upper bound for this probability.}

\subsection{Justification of Sparsity Assumption}\label{sparsity_sec}

\begin{figure}[htbp]
\centering
\includegraphics[width=2.7 in]{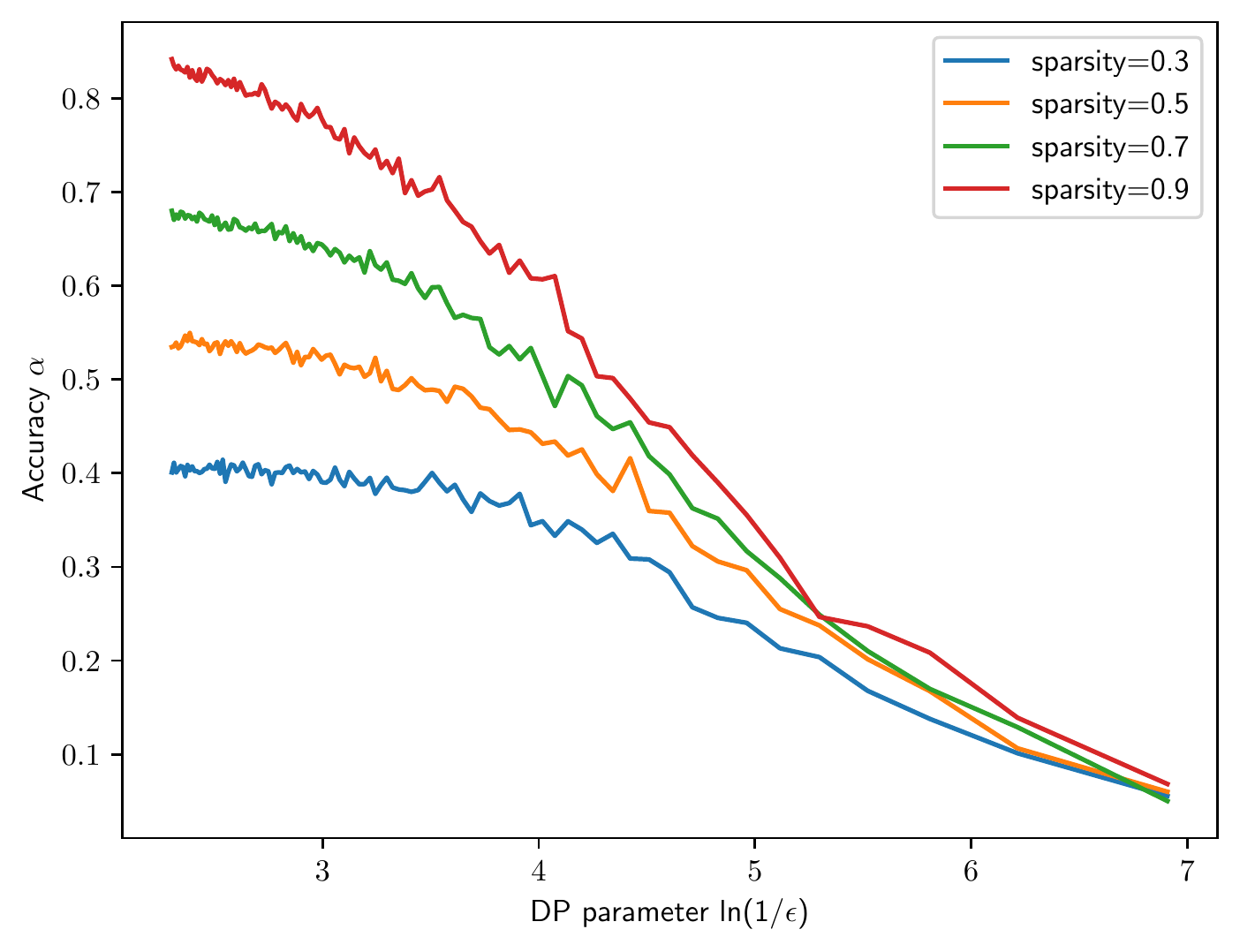}\vspace{-0.2cm}
\caption{Performance of One-shot compressive sensing inference for different sparsity}
\label{spar}
\end{figure}

{
While it is hard to theoretically examine how the failure to meet this assumption may affect the performance of the proposed inference framework, we provide the following numerical analysis.
Define the degree of sparsity $s$ for vector $X \in \{0,1\}^{N \times T}$ as follows:
\begin{equation}
    s = 1-\frac{\sum_{t=1}^{T-1}\|X_{t+1}-X_{t}\|_0}{N(T-1)}.
\end{equation}
This metric allows us to synthetic load data with different sparsity levels to our theoretical conclusions. Fig. \ref{spar} plots the relationship between DP parameter and inference accuracy for different sparsity levels, from $0.3$ to $0.9$. It is clear that the general trends remain the same for different sparsity levels while higher sparsity level leads to better performance.
}

{
We further illustrate the sparsity levels of three most widely adopted public datasets in Table \ref{sptable}. The sparsity levels of these three datasets are all very high.}

\begin{table}[ht!]
\caption{The Sparsity of Three Widely Adopted Public Datasets}
\label{sptable}
\centering
\begin{tabular}{l|l}
\hline
Dataset & Sparsity \\ \hline
UK-DALE & 99.86\%  \\ \hline
TEALD   & 99.29\%  \\ \hline
REDD    & 98.97\%  \\ \hline
\end{tabular}
\end{table}

\subsection{Proof of Lemma 1}\label{Lemma 1}

Denote the optimal solution to (P1) by 
$\Delta_{t}^{0}$ and the optimal solution to (P2)
by $\Delta_{t}$. By the equivalence of optimal solution set, we mean
\begin{equation}
\| \Delta_t^0 \|_0 = \| \Delta_t \|_0.
\end{equation}

Hence, it suffices for us to show two cases: $\| \Delta_t^0 \|_0 \le \| \Delta_t \|_0$ and $\| \Delta_t^0 \|_0 \ge \| \Delta_t \|_0$. Note that, it is straightforward to see that all feasible solutions to (P1) are also feasible to (P2). Hence, the second inequality immediately follows. All that remains is to prove the first inequality.

We prove it by contradiction by utilizing the crucial technical alignment assumption: Assumption 3. It states that for the ascending order of $\mathbf{P}$'s 
entries $\{P^\epsilon_{1},..,P^\epsilon_{N}\}$, and for all 
$U \leq U_{t}$, it holds:
\begin{equation}
\begin{aligned}
\sum\nolimits_{k=1}^{U}P_{k}^\epsilon-\sum\nolimits_{k=1}^{U-1}P^\epsilon_{N+1-k}>2\delta.
\end{aligned}    
\end{equation}

Substituting the sparsity assumption into the sensitivity constraint yields that there exists $P_1$,..,$P_U$ satisfying:
\begin{equation}
\begin{aligned}\label{ass_lemma1}
\sum\nolimits_{k=1}^{U}P_{i}\leq K_{t}+\delta.
\end{aligned}    
\end{equation}

It holds for all possible $\underline{U}<U$,
\begin{equation}\label{eq2_lemma1}
\begin{aligned}
&\sum\nolimits_{k=1}^{\underline{U}}P_{N+1-k}^{\epsilon}
< \sum\nolimits_{k=1}^{\underline{U}}P_{i}-2\delta\\
< &\sum\nolimits_{k=1}^{\underline{U}}P_{k}^\epsilon-2\delta \leq K_{t}-\delta.
\end{aligned}    
\end{equation}

Note that, Eqs. (\ref{ass_lemma1}) and (\ref{eq2_lemma1}) hold for both problems (P1) and (P2). 

Denote $\| \Delta_t^0 \|_0$ by $U$, and denote $\| \Delta_t \|_0$ by $\underline{U}$. Suppose $\underline{U}<U$, Eqs. (\ref{ass_lemma1}) and (\ref{eq2_lemma1}) yield that
\begin{equation}
\begin{aligned}
\Delta_t\mathbf{P}\leq\sum\nolimits_{k=1}^{\underline{U}}P_{N+1-k}^\epsilon <K_{t}-\delta,
\end{aligned}    
\end{equation}
This indicates that $\Delta_t$ does not satisfy the sensitivity constraint in (P2), which contradicts to our assumption that it is the optimal solution to (P2). Thus, we could know that $\| \Delta_t^0 \|_0 \le \| \Delta_t \|_0$. This concludes our proof.

\subsection{Proof for Theorem 1}\label{Theorem 1}
We first define the notations and present the lemma that we use {to prove} this theorem.

For a set $U\subseteq \{1,\cdots,N\}$, define 
vector $\mathbf{P}_U$ as follows:
\begin{equation}
    \mathbf{P}_U=\left\{P_i\bigg|P_i\in \mathbf{P},i\in U\right\}.
\end{equation}
We say $\mathbf{P}_U$ is $S$-bounded by $\delta_S$, if the following holds for all $c$, and for all $U$ such that $|U|\le S$:
\begin{equation}
    (1-\delta_{S})\|c\|_{2}^{2}\leq\|\mathbf{P}_{U}\cdot c\|_{2}^2\leq(1+\delta_{S})\|c\|_{2}^{2}.
\end{equation}

For our problem, the sparsity assumption implies that $\Delta_t$ has at most $U_t$ non-zero entries. Assume $\mathbf{P}_U$ is both $3U_t$-bounded and $4U_t$-bounded, and the 
corresponding bounds are $\delta_{3U_t}$ and 
$\delta_{4U_t}$, respectively. We use Theorem 1 
in \cite{candes2006stable} as Lemma 2 
for our subsequent proof.

\vspace{0.1cm}
\begin{lemma}\label{lemma2_}\rm
(\noindent $\textbf{Theorem 1}$ in \cite{candes2006stable}):
For $X_{t}$ with $U_t$ non-zero entries, and for the matrix $P$ with $n$ columns, if $P$ is 
both $3U_t$-bounded and $4U_t$-bounded, the difference between the solution for (P3) and ground truth is upper bounded, i.e.,
\begin{equation}\label{Lemma2_delta}
   \|\Delta_{t}^{*}-\Delta_{t}^{0}\|_{2}\leq \frac{4}{(\sqrt{3(1-\delta_{4U_t})}-\sqrt{1+\delta_{3U_t}})}\cdot\delta.
\end{equation}
\end{lemma}

To apply Lemma 2 to our problem, we need to express the constraint $\|\Delta_t\boldsymbol{P} - K_t\|_2 < \delta$ in the following equivalent form:
\begin{equation}
\label{Lemma2_delta2}
\left|\left|\Delta_{t}\frac{\mathbf{P}}{\|\mathbf{P}\|_{2}}-\frac{K_{t}}{\|\mathbf{P}\|_{2}}\right|\right|_{2}<\frac{\delta}{\|\mathbf{P}\|_{2}}.
\end{equation}

Thus, Lemma 2 directly dictates
\begin{equation}\label{Lemma2_delta3}
    \|\Delta_{t}^{*}-\Delta_{t}^{0}\|_{2}\leq C(\mathbf{P})\cdot\delta,
\end{equation}
where
\begin{equation}\label{Lemma2_delta4}
    C(\mathbf{P})=\frac{4}{(\sqrt{3(1-\delta_{4U_t})}-\sqrt{1+\delta_{3U_t}})\|\mathbf{P}\|_{2}}.
\end{equation}

Note that $\mathbf{E}[\bar{\Delta}_t]=\Delta_{t}^{*}$, which yields
\begin{equation}\label{pn}
\begin{aligned}
\|\mathbf{E}[\bar{\Delta}_t]-\Delta_{t}^{0}\|_{2}&\leq \|\mathbf{E}[\bar{\Delta}_t]-\Delta_{t}^{*}\|_{2}+\|\Delta_{t}^{*}-\Delta_{t}^{0}\|_{2}\\
&\leq C(\mathbf{P})\cdot\delta.  
\end{aligned}
\end{equation}
This completes our proof for Theorem 1.

\subsection{Proof for Proposition 1}\label{Proposition 1}

Note that the following triangle inequality holds with $K_{t}^\epsilon$ defined in (P4):
\begin{equation}\label{triangle}
\begin{aligned}
&|\Delta_{t}^\epsilon\mathbf{P}-\Delta_{t}^{0}\mathbf{P}|\\
\leq&|\Delta^\epsilon_t\mathbf{P}-\Delta^*_t\mathbf{P}+\Delta^*_t\mathbf{P}-\Delta^0_t\mathbf{P}|\\
\leq&|\Delta^\epsilon_t\mathbf{P}-\Delta^*_t\mathbf{P}|+|\Delta^*_t\mathbf{P}-\Delta^0_t\mathbf{P}|\\
\leq&C(\mathbf{P})\cdot\delta+|\Delta^\epsilon_t-K_t^\epsilon+K_t^\epsilon-\Delta^*_t\mathbf{P}|\\
\leq&C(\mathbf{P})\cdot\delta+\delta+||y_t-y_{t+1}+n_t-n_{t+1}|-\Delta^*_t\mathbf{P}|\\
\leq&C(\mathbf{P})\cdot\delta+\delta+||y_t-y_{t+1}|+|n_t-n_{t+1}|-\Delta^*_t\mathbf{P}|\\
\leq&C(\mathbf{P})\cdot\delta+\delta+\delta+|n_t-n_{t+1}|\\
\leq&(C(\mathbf{P})+2)\cdot\delta+|n_t-n_{t+1}|
\end{aligned}
\end{equation}
{Proposition 1 immediately follows.}

\subsection{Proof for Theorem 3}\label{Theorem 3}

\vspace{0.1cm}
Denote $l_{t}=n_{t}-n_{t-1}$. Since $n_{t}$ and $n_{t-1}$ are \emph{i.i.d} Laplace noises, we can show that 
\begin{equation}
\begin{aligned}
\mathbf{E}[|l_{t}|]=2\int^{\infty}_{0}\!\!x\left(\frac{x}{4\lambda^{2}}\!+\!\frac{1}{4\lambda}\right)e^{-\frac{x}{\lambda}}dx
=\frac{3\lambda}{2}
=\frac{3\delta}{4\epsilon}.
\end{aligned}
\end{equation}
Note that the last equality holds because $\Delta{f} = \frac{\delta}{2}$ and $\lambda = \frac{\Delta{f}}{\epsilon}$.

This allows us to express $\mathbf{E}[\alpha]$ in terms of $\delta$ and $\epsilon$:

\begin{equation}\label{accp}
\begin{aligned}
&\mathbf{E}[\alpha]=\mathbf{E}_{l_{t}}\left[1-\frac{\mathbf{E}\|\bar{\Delta}_t^\epsilon-\Delta_{t}^{0}\|_{1}}{N}\right]\\
=& \mathbf{E}_{l_{t}}\left[1\!-\!\frac{\mathbf{E}\|\bar{\Delta}_t^\epsilon\!-\!\Delta_{t}^{0}\|_{2}}{N}\right]\!=\!\mathbf{E}_{l_{t}}\left[1\!-\!\frac{\|\mathbf{E}[\bar{\Delta}_t^\epsilon]\!-\!\Delta_{t}^{0}\|_{2}}{N}\right]\\
\geq&\ \mathbf{E}_{l_{t}}\left[\max(1-\frac{(C(\mathbf{P})+2)\delta+|l_{t}|}{N},0)\right]. 
\end{aligned}
\end{equation}
Expressing the expectation in the integral form further yields that

\begin{equation}
\begin{aligned}
&\mathbf{E}[\alpha]\\
&\geq1-2\int_{N-(C(\mathbf{P})+2)\delta}^{\infty}\!\left(\frac{x}{4\lambda^{2}}\!+\!\frac{1}{4\lambda}\right)e^{-\frac{x}{\lambda}}dx\\
&-2\int_{0}^{N-(C(\mathbf{P})+2)\delta}\!\left(\frac{(C(\mathbf{P})+2)\delta+x}{N}\right)\left(\frac{x}{4\lambda^{2}}\!+\!\frac{1}{4\lambda}\right)e^{-\frac{x}{\lambda}}dx.
\end{aligned}
\end{equation}
From Eq. (\ref{accp}), the inequality holds due to the fact that accuracy has a trivial lower bound 0, which indicates the error has a trivial upper bound 1. Theorem 3 follows by further standard mathematical manipulations.

\subsection{Proof for Theorem 4}\label{Theorem 4}
The accuracy is defined as follows 
\begin{equation}
\begin{aligned}
\alpha = 1 - \frac{\mathbf{E}\|\bar{\Delta_{t}^\epsilon}-\Delta_{t}^{0}\|_{1}}{N}.
\end{aligned}
\end{equation}
To find its upper bound, we need to find the lower bound of the error.
{With $K_{t}^\epsilon$ in Problem (P4),} we define event $\Omega$ as $\{K_{t}^\epsilon \not \in [0,\Delta^0_t\mathbf{P}+2\delta]\}\cap\{(y_t-y_{t+1})(n_t-n_{t+1})>0\}$.

This event $\Omega$ allows us to lower bound the error:
\begin{equation}
\begin{aligned}
\label{omegaerr}
&\mathbf{E}[1-\alpha] =\mathbf{E}_{K_{t}^\epsilon}\frac{\mathbf{E}\|\bar{\Delta}_{t}^\epsilon-\Delta_{t}^{0}\|_{1}}{N} \\
=&\mathbf{E}_{K_{t}^\epsilon}\frac{\|\mathbf{E}[\bar{\Delta}_t^\epsilon]-\Delta_{t}^{0}\|_{2}}{N}\geq \mathbf{E}\left[\frac{\|\mathbf{E}[\bar{\Delta}_t^\epsilon ]-\Delta_{t}^{0}\|_{2}}{N}\mathbf{I}\{\Omega\}\right],
\end{aligned}
\end{equation}
where $\mathbf{I}\{\cdot\}$ is the indicator function.

For a certain $K_t^\epsilon$, from the definition of rounding, we know that the expectation of rounding is the same as the optimal solution to (P3), denoted by $\Delta_{t}^{*}$.
Hence, we could substitute $\mathbf{E}[\bar{\Delta}_t^\epsilon]$ with $\Delta_{t}^{*}$.

When the event $\mathscr{\Omega}$ happens,  $K_{t}^\epsilon$ could be related to the ground truth $\Delta_{t}^{0}$, in which $l_{t}$ denotes the $n_{t}-n_{t-1}$.
Since the $(y_t-y_{t+1})(n_t-n_{t+1})>0$, we could derive that, 
\begin{equation}
\begin{aligned}\label{kt}
K_{t}^\epsilon=|y_{t}-y_{t-1}|+|l_{t}|
&\geq\Delta_{t}^{0}P-\delta+|l_{t}|.
\end{aligned}
\end{equation}
Thus, if the $\Delta_{t}^{*}$ is feasible, i.e.,
\begin{equation}
\begin{aligned}
|\Delta_{t}^{*}\mathbf{P}-K_{t}^\epsilon|\le \delta,
\end{aligned}
\end{equation}
then we know
\begin{equation}\label{alx_2}
\begin{aligned}
\Delta_{t}^{*}\mathbf{P}-K_{t}^\epsilon\ge -\delta.
\end{aligned}
\end{equation}

When the event $\mathscr{\Omega}$ happens, we could conclude that $|l_{t}|>2\delta$. 
Therefore, combining with Eqs.(\ref{kt}) and (\ref{alx_2}) yields
\begin{equation}
\begin{aligned}
||l_{t}|-2\delta|&\leq|(\Delta_{t}^{*}-\Delta_{t}^{0})\mathbf{P}|\\
&\leq \|\Delta_{t}^{*}-\Delta_{t}^{0}\|_{2}\|\mathbf{P}\|_{2}.
\end{aligned}
\label{l_delta}
\end{equation}

{Plugging} Eq. (\ref{l_delta}) into Eq. (\ref{omegaerr}) further implies that
\begin{equation}
\begin{aligned}
\mathbf{E}[1-\alpha]
\geq &\mathbf{E}\left[\frac{\|(\mathbf{E}[\bar{\Delta}_t^\epsilon ]-\Delta_{t}^{0})\|_{2}}{N}\mathbf{I}\{\Omega\}\right]
\\
\geq &\mathbf{E}\left[\frac{||l_{t}|-2\delta|}{N\|\mathbf{P}\|_{2}}\mathbf{I}\{\Omega\}\right]\\
= &\int^{2n\|\mathbf{P}\|_{2}+2\delta}_{2\delta}\frac{||x|-2\delta|}{N\|\mathbf{P}\|_{2}}(\frac{x}{4\lambda^{2}}+\frac{1}{4\lambda})e^{-\frac{x}{\lambda}}dx\\
&+\int_{\infty}^{2n\|\mathbf{P}\|_{2}+2\delta}(\frac{x}{4\lambda^{2}}+\frac{1}{4\lambda})e^{-\frac{x}{\lambda}}dx.\\
\end{aligned}
\end{equation}

The inequalities are {again} making use of the fact that accuracy has a trivial lower bound $0$. With mathematical manipulations, we can prove Theorem 4.

\subsection{Proof for Theorem 5}\label{Theorem 5}
For one-shot compressive sensing, with the above proofs of Theorem 3 and 4, we could derive the bounds for the difference between the solution to problem (P4) $\Delta_t^{*}$ and the ground truth $\Delta_t^{0}$ {in the following form}:
\begin{equation}
g(\delta,\epsilon)\leq\mathbf{E}\|\Delta_t^{*}-\Delta_t^{0}\|_2\leq G(\delta,\epsilon).
\end{equation}
{Next, we construct $g(\delta,\epsilon)$ and $G(\delta,\epsilon)$. Following the rounding procedure,} for the appliance $k$, we denote its approximate and true states as $X_t^{k}$ and $X_t^{0k}$, respectively. We also denote its approximate and true switching events by $\Delta_t^{*k}$ and $\Delta_t^{0k}$. The procedure leads to the following equations:
\begin{align}
X_{t+1}^{k} &= \Delta_t^{*k}(1-X_t^{k})+(1-\Delta_t^{*k})X_t^{k},\\
X_{t+1}^{0k} &= \Delta_t^{0k}(1-X_t^{0k})+(1-\Delta_t^{0k})X_t^{0k}.
\end{align}

Together, they indicate
\begin{equation}
\begin{aligned}\label{res}
&|X_{t+1}^{k}-X_{t+1}^{0k}|\\
=&|X_{t}^{k}-X_{t}^{0k}+\Delta_t^{*k}-\Delta_t^{0k}+2(X_{t}^{0k}\Delta_t^{0k}-X_{t}^{k}\Delta_t^{*k})|\\
=&|(1-2\Delta_t^{*k})(X_{t}^{k}-X_{t}^{0k})+(1-2X_{t}^{0k})(\Delta_t^{*k}-\Delta_t^{0k})|\\
\leq& |(1-2\Delta_t^{*k})(X_{t}^{k}-X_{t}^{0k})|+|(1-2X_{t}^{0k})(\Delta_t^{*k}-\Delta_t^{0k})|\\
\leq&|X_{t}^{k}-X_{t}^{0k}|+|\Delta_t^{*k}-\Delta_t^{0k}|.
\end{aligned}
\end{equation}

The last inequality holds because either $\Delta_t^{*k}$
or $X_{t}^{0k}$ is in $[0,1]$. Hence, both $1-2\Delta_t^{*k}$
and $1-2X_{t}^{0k}$ are in $[-1,1]$.

Taking the expectation, we have
\begin{equation}
\begin{aligned}
&\mathbf{E}|X_{t}^{k}-X_t^{0k}|\\
\leq&\mathbf{E}\sum\nolimits_{i=1}^{T-1}|\Delta_i^{*k}-\Delta_i^{0k}| =\mathbf{E}\sum\nolimits_{i=1}^{T-1}\|\Delta_i^{*k}-\Delta_i^{0k}\|_{2}\\
\leq &\mathbf{E}\sum\nolimits_{i=1}^{T-1}\|\Delta_i^{*}-\Delta_i^{0}\|_{2}
=(T-1)G(\delta,\epsilon).
\end{aligned}
\end{equation}

Denoting $\boldsymbol{X}^{0}$ as the ground truth and $\boldsymbol{X}$ as the results before rounding, we know
\begin{equation}
\begin{aligned}
&\mathbf{E}\|X-X^{0}\|_{1}\\
=&\mathbf{E}\sum_{k=1}^{N}\sum_{i=1}^{T}|X_{i}^{k}-X_{i}^{0k}|\leq\mathbf{E}\frac{T(T-1)}{2}G(\delta,\epsilon)N.
\end{aligned}
\end{equation}

Then, back to {lower bound} the accuracy, in which $\bar{\boldsymbol{X}^{\epsilon}}$ denotes the actual decoding results after rounding, we could derive that:
\begin{equation}
\begin{aligned}
\mathbf{E}[\alpha_m] &= 1 - \frac{\mathbf{E}\|\bar{\boldsymbol{X}^{\epsilon}}-\boldsymbol{X}^{0}\|_{1}}{NT} = 1 - \frac{\mathbf{E}\|\mathbf{E}[\bar{\boldsymbol{X}^{\epsilon}}]-\boldsymbol{X}^{0}\|_{1}}{NT}\\
&=1 - \frac{\|\boldsymbol{X}-\boldsymbol{X}^{0}\|_{1}}{NT}\geq 1-\frac{(T-1)}{2}G(\delta,\epsilon).
\end{aligned}
\end{equation}

We construct $G(\delta,\epsilon)$ in terms of the accuracy bound for the one-shot algorithm. 
If we denote the lower and upper bounds for one-shot compressive sensing algorithm by $b(\delta,\epsilon)$ and $B(\delta,\epsilon)$,
then from Eq. (\ref{accp}), we have
\begin{equation}
\begin{aligned}
\mathbf{E}\left[1-\frac{\|\mathbf{E}[\bar{\Delta}_t^\epsilon]-\Delta_{t}^{0}\|_{2}}{N}\right]&=\mathbf{E}\left[1-\frac{\|\Delta_{t}^{*}-\Delta_{t}^{0}\|_{2}}{N}\right]\\
&=\mathbf{E}[\alpha]
\geq b(\delta,\epsilon).
\end{aligned}
\end{equation}
This constructs the lower bound:
\begin{equation}\label{ulbound_trans}
G(\delta,\epsilon)=1-b(\delta,\epsilon)N.
\end{equation}
Specifically, we could derive the lower bound for multi-shot compressive sensing as

\begin{equation}
\mathbf{E}[\alpha_m]\geq 1-\frac{(T-1)(1-b(\delta,\epsilon)N)}{2}.   
\end{equation}

As for the upper bound, we could construct an oracle to analyze a simple bound. The oracle could identify all the true states when $t>1$. When $t=1$, it does the same as our 
multi-shot compressive sensing algorithm. 
Hence, its error totally comes from the error for time period $1$. Since we know the initial states, our error comes from the first period's one-shot compressive sensing's calculations,
that is the error for $\bar{\Delta}_1^\epsilon$. This observation yields that
\begin{equation}
\mathbf{E}\left[1-\frac{\|\mathbf{E}[\bar{\Delta}_t^\epsilon]-\Delta_{t}^{0}\|_{2}}{N}\right]\leq B(\delta,\epsilon).
\end{equation}
This further implies that
\begin{equation}
\mathbf{E}\left[\frac{\|\mathbf{E}[\bar{\Delta}_t^\epsilon]-\Delta_{t}^{0}\|_{2}}{N}\right]\geq 1-B(\delta,\epsilon).
\end{equation}
Therefore, the upper bound could be derived with the help of Theorem 4 as follows:
\begin{equation}
\begin{aligned}
\mathbf{E}[\alpha_m]&=1-\frac{\mathbf{E}\|\bar{\boldsymbol{X}^{\epsilon}}-\boldsymbol{X}^{0}\|_{1}}{NT}\leq1 - \frac{\mathbf{E}\|\bar{X_1^{\epsilon}}-X_1^{0}\|_{1}}{NT}\\
&=
\mathbf{E}\left[1-\frac{\|\mathbf{E}[\bar{\Delta}_t^\epsilon]-\Delta_{t}^{0}\|_{2}}{NT}\right]\leq 1-\frac{1-B(\delta,\epsilon)}{T}.
\end{aligned}
\end{equation}
Thus, we complete our proof.

\subsection{Proof for Proposition 2}\label{Proposition 2}

We first define the notion of a good hierarchy. For a hierarchy $\mathscr{H}$ with $S$ items, we rank the items ascendingly: $P_{1},..,P_{S}$. If $\mathscr{H}$ satisfies that
\begin{equation}\label{pro2us}
\sum\nolimits_{i=1}^{U}P_{i}-2\delta>\sum\nolimits_{i=1}^{U-1}P_{S-i+1}, \forall U<S,
\end{equation}
we call it a good hierarchy. Notice this characteristic is stronger than Assumption 3. For a good hierarchy with $S$ items, if $S\leq U_t$, Assumption 3 immediately holds.

We prove the proposition by induction. The induction basis is clear: the single item hierarchy is a trivial good hierarchy. The key step is the induction process, where we want to  show that for a good hierarchy with $S$ items, if the $(S+1)^{th}$ appliance satisfies,

\begin{equation}\label{decondition}
\sum\nolimits_{i=1}^{\lfloor\frac{S}{2}\rfloor+1}P_{i}-2\delta\geq\sum\nolimits_{i=1}^{\lfloor\frac{S}{2}\rfloor-1}P_{S+1-i}+P_{S+1},
\end{equation}
then we should {include} it to the set and form a larger good hierarchy. This can be proved by examining the following two cases. 

For $U\leq \lfloor\frac{S}{2}\rfloor+1$, since
\begin{equation}
\sum\nolimits_{i=U+1}^{\lfloor\frac{S}{2}\rfloor+1}P_{i}<\sum\nolimits_{i=U-1}^{\lfloor\frac{S}{2}\rfloor-1}P_{S-i+1},
\end{equation}
together with condition (\ref{decondition}), we could obtain that
\begin{equation}
\sum\nolimits_{i=1}^{U}P_{i}-2\delta>\sum\nolimits_{i=1}^{U-2}P_{S-i+1}+P_{S+1}.
\end{equation}

For $\lfloor\frac{S}{2}\rfloor+1<U\le S$, we can observe the following two facts:

\begin{equation}
\sum\nolimits_{S-U+3}^{U}P_{i}=\sum\nolimits_{i=S-U+1}^{U-2}P_{S-i+1},
\end{equation}
and
\begin{equation}
    S-U+2\leq\lfloor\frac{S}{2}\rfloor+1.
\end{equation}

Together, they imply that
\begin{equation}
\sum\nolimits_{i=1}^{S-U+2}P_{i}-2\delta>\sum\nolimits_{i=1}^{S-U}P_{S-i+1}+P_{S+1}.
\end{equation}
Combining the two cases, we can conclude the induction. Following the same routine, we can show that, condition (\ref{decondition}) is also necessary. That is, if the $(S+1)^{th}$ appliance does not satisfy this condition, it should not be {included} into the current set (hierarchy), as it will make the hierarchy not good. We should, instead, start a new set. Thus, we complete the proof for Proposition 2.

\subsection{Proof for Theorem 6}\label{Theorem 6}

We prove the following Lemma, which is crucial to our proof.

\vspace{0.1cm}
\begin{lemma}\label{lemma3_}\rm
For our one-shot compressive sensing algorithm, we denote the lower and upper bound as $b(\delta,\epsilon)$ and $B(\delta,\epsilon)$. Besides the Laplace noises, we inject new disturbances $\xi_t$ and $\xi_{t+1}$ into $y_t+n_t$ and $y_{t+1}+n_{t+1}$, in which $y_t$ and $y_{t+1}$ is the meter readings and $n_t$ is the injecting Laplace noise as we defined in Eq. (\ref{ps}). If the disturbances satisfy that
$\mathbf{E}|\xi_t-\xi_{t+1}|\leq Y$, then the bounds associated with our algorithm with disturbances are $b(\delta+\frac{Y}{2+C(\mathbf{P})},\epsilon)$ and $B(\delta+\frac{Y}{2},\epsilon)$, respectively.
\end{lemma}
\vspace{0.1cm}

\noindent $\textbf{Proof}$: 
First we construct the new lower bound. Following the same routine as Eq. (\ref{triangle}), {we have}

\begin{equation}\label{lboundextension}
\begin{aligned}
&|\Delta_{t}^\epsilon\mathbf{P}-\Delta_{t}^{0}\mathbf{P}|\\
\leq& (C(\mathbf{P})+2)\cdot\delta+|n_{t}-n_{t+1}+\xi_t-\xi_{t+1}|.\\
\leq & (C(\mathbf{P})+2)\cdot\delta+|n_{t}-n_{t+1}|+|\xi_t-\xi_{t+1}|.
\end{aligned}
\end{equation}

This implies that
\begin{equation}
\begin{aligned}
\mathbf{E}[\alpha]&\geq\mathbf{E}\left[1-\frac{(C(\mathbf{P})+2)\delta+|\xi_t-\xi_{t+1}|+|l_{t}|}{N}\right]\\
&\geq\mathbf{E}\left[1-\frac{(C(\mathbf{P})+2)\delta+Y+|l_{t}|}{N}\right]\\
&=\mathbf{E}\left[1-\frac{(C(\mathbf{P})+2)(\delta+\frac{Y}{2+C(\mathbf{P})})+|l_{t}|}{N}\right],
\end{aligned}
\end{equation}
where $l_t$ is the difference between the Laplace noises, i.e., $l_t=n_t-n_{t+1}$. Clearly, the disturbance plays a similar role as the noises. Revisiting Eq. (\ref{kt}) yields that

\begin{equation}
K_{t}^\epsilon=|y_{t}-y_{t+1}+\xi_{t}-\xi_{t+1}|+|l_{t}|.
\end{equation}

Hence, we could establish the new upper bound as follows:
\begin{equation}
\begin{aligned}
&||l_{t}|-2\delta-Y|\leq|\mathbf{E}_{\xi}[|l_{t}|-2\delta-|\xi_{t}-\xi_{t+1}|]|\\
\leq &\mathbf{E}_{\xi}||l_{t}|-2\delta-|\xi_{t}-\xi_{t+1}||\\
\leq&\mathbf{E}_{\xi}|(\Delta_{t}^{*}-\Delta_{t}^{0})\mathbf{P}|
\leq \mathbf{E}_{\xi}\|\Delta_{t}^{*}-\Delta_{t}^{0}\|_{2}\|\mathbf{P}\|_{2}.
\end{aligned}
\end{equation}

The upper bound can be derived following the same routine. Then we finish our proof of Lemma 3.

\vspace{0.3cm}
The key to apply Lemma 3 to the proof of Theorem 6 is to characterize the disturbance connections between hierarchies. We characterize such connections by induction. The induction basis is to examine hierarchy 1, where the disturbances come from all the other hierarchies' instead of decoding error.

Specifically, for $\forall y_t,y_{t+1}$, if we denote $y^{01}_t$ as the hierarchy $1$'s meter reading at time $t$ and $y^{out_1}_t$ as the other hierarchies meter reading at time $t$, it holds that:
\begin{equation}
|y_t-y_{t+1}|=|y^{01}_{t}-y^{01}_{t+1}+l_t+y^{out_1}_{t}-y^{out_1}_{t+1}|.
\end{equation}
With the sparsity $U$, we could further derive that 
\begin{equation}
|y^{out_1}_t-y^{out_1}_{t+1}|\leq |P_{U}^1|,
\end{equation}
where $P_{U}^1$ is defined as the maximum summation of $U$ appliances whose power is smaller than the smallest energy consumption in the hierarchy $1$, which is denoted {by} $P_{m}^1$. Note that $\xi_{t}=y^{out_1}_{t}$, i.e., 

\begin{equation}
|\xi_t-\xi_{t+1}|\leq P_{U}^1.
\end{equation}
This constructs our induction basis with the above Lemma 3.

Next, we analyze the bounds for hierarchy $i$. We assume for each hierarchy $k$, $1\le k<i$, we have

\begin{equation}
    m_{k}\leq\mathbf{E}[\alpha_{k}]\leq M_{k},
\end{equation}
where $m_k$ is the lower bound of the multi-shot compressive sensing algorithm accuracy for hierarchy $k$ and $M_k$ is the upper bound for this hierarchy as indicated in Theorem 6. 

Thus, denote the hierarchy $i$'s meter reading as $y^{0i}_t$ at time $t$ and the decoding results as $\tilde{y}^{i}_t$. By the definition of $\alpha_k$, we can show that $\forall k<i$,
\begin{equation}
\begin{aligned}
\mathbf{E}|\tilde{y}^{k}_t-y_t^{0k}|&=\mathbf{E}|(\bar{\boldsymbol{X}^{\epsilon}[k]}-\boldsymbol{X}^{0}[k])\mathbf{P}_k|\\
&\leq\mathbf{E}\|(\bar{\boldsymbol{X}^{\epsilon}[k]}-\boldsymbol{X}^{0}[k])\|_2\|\mathbf{P}_k\|_2\\
&=N_{k}T(1-\mathbf{E}[\alpha_i])\|\mathbf{P}_k\|_2\\
&\leq N_{k}T(1-m_{k})\|\mathbf{P}_k\|_2.
\end{aligned}
\end{equation}

Also, if we denote all hierarchies whose index $m>i$'s actual total meter reading as $y^{out_i}_t$,
it could be directly derived from the sparsity of switching events that:
\begin{equation}
|y^{out_i}_t-y^{out_i}_{t+1}|\leq |P_U^i|
\end{equation}

Thus, the total disturbance for hierarchy $i$ can be calculated as follows:
\begin{equation}
\xi_t = \sum\nolimits_{k=1}^{i-1}(y^{0k}_t-\tilde{y}^{k}_t)+y^{out_i}_t.
\end{equation}
This allows as to construct the bounds for $\mathbf{E}|\xi_t-\xi_{t+1}|$:
\begin{equation}
\begin{aligned}
&\mathbf{E}|\xi_t-\xi_{t+1}|\\
\leq &\mathbf{E}[\sum_{k=1}^{i-1}(|y^{0k}_t-\tilde{y}^{k}_t|+|y^{0k}_{t+1}-\tilde{y}^{k}_{t+1}|)+|y^{out_i}_t-y^{out_i}_{t+1}|]\\
\leq& 2\sum\nolimits_{k=1}^{i-1}N_{k}T(1-m_{k})\|\mathbf{P}_k\|_2+|P_U^i|.
\end{aligned}
\end{equation}
Together with the Lemma 3, we complete the induction, and hence prove Theorem 6.

\end{document}